\documentclass[10pt,journal,compsoc]{IEEEtran}

\ifCLASSOPTIONcompsoc
  \usepackage[nocompress]{cite}
\else
  \usepackage{cite}
\fi
\ifCLASSINFOpdf
  \usepackage[pdftex]{graphicx}
\else
\fi
\graphicspath{{figs/}{figures/}{pictures/}{images/}{./}} 
\usepackage{balance}

\usepackage{url}
\usepackage{hyperref}

\usepackage{wrapfig}
\usepackage{threeparttable}
\usepackage{caption}
\usepackage{subcaption}

\usepackage{mfirstuc}

\usepackage{amsmath}
\usepackage[capitalise]{cleveref}
\usepackage{enumitem}
\usepackage{paralist}
\usepackage{multirow}
\usepackage{makecell}
\usepackage{algorithm} 
\usepackage{algorithmic}  
\usepackage[algo2e]{algorithm2e} 

\usepackage{dblfloatfix}
\usepackage{balance}
\usepackage[dvipsnames]{xcolor}
\newcommand{\prefix}{\textcolor[RGB]{0, 102, 204}}
\newcommand{\name}{{\textit{MARLens}}}

\newcommand{\controlpanelCapital}{{\textcolor{black}{Control Panel}}}

\newcommand{\trainingdistribution}{{\textcolor{black}{training distribution}}}
\newcommand{\trainingdistributionCapital}{{\textcolor{black}{Training Distribution}}}

\newcommand{\episodeoverviewCapital}{{\textcolor{black}{Episode Overview}}}

\newcommand{\episodedetailCapital}{{\textcolor{black}{Episode Detail}}}

\newcommand{\policyexplainerCapital}{{\textcolor{black}{Policy Explainer}}}

\newcommand{\simulationreplayCapital}{{\textcolor{black}{Simulation Replay}}}

\newcommand{\snapshotlogCapital}{{\textcolor{black}{Snapshot Log}}}

\newcommand{\zyt}{\textcolor{black}}

\newcommand{\revisedOld}{\textcolor{black}}
\newcommand{\revised}{\textcolor{black}}

\usepackage{tikz}

\begin{document}
%
\title{\name: Understanding Multi-agent Reinforcement Learning for Traffic Signal Control via Visual Analytics}
%
%
%
%

\author{Yutian Zhang, Guohong Zheng, Zhiyuan Liu, Quan Li and Haipeng Zeng
\IEEEcompsocitemizethanks{
\IEEEcompsocthanksitem Y. Zhang, G. Zheng, Z. Liu are with Sun Yat-sen University. E-mail: \{zhangyt85, zhenggh8, liuzhy253\}@mail2.sysu.edu.cn.

\IEEEcompsocthanksitem Q. Li is with the School of Information Science and Technology, ShanghaiTech University. E-mail: liquan@shanghaitech.edu.cn.

\IEEEcompsocthanksitem H. Zeng is with Sun Yat-sen University and is the corresponding author. E-mail: zenghp5@mail.sysu.edu.cn.

}
\thanks{Manuscript received March 31, 2023; revised December 07, 2023.}}

%
%

\markboth{IEEE TRANSACTIONS ON VISUALIZATION AND COMPUTER GRAPHICS}%
{Shell \MakeLowercase{\textit{et al.}}: Bare Advanced Demo of IEEEtran.cls for IEEE Computer Society Journals}
%



\IEEEtitleabstractindextext{%
\begin{abstract}
\revisedOld{The issue of traffic congestion poses a significant obstacle to the development of global cities. One promising solution to tackle this problem is intelligent traffic signal control (TSC). Recently, TSC strategies leveraging reinforcement learning (RL) have garnered attention among researchers. However, the evaluation of these models has primarily relied on  fixed metrics like reward and queue length. This limited evaluation approach provides only a narrow view of the model's decision-making process, impeding its practical implementation. Moreover, effective TSC necessitates coordinated actions across multiple intersections. Existing visual analysis solutions fall short when applied in multi-agent settings. In this study, we delve into the challenge of interpretability in multi-agent reinforcement learning (MARL), particularly within the context of TSC. We propose {\name}, a visual analytics system tailored to understand MARL-based TSC. Our system serves as a versatile platform for both RL and TSC researchers. It empowers them to explore the model's features from various perspectives, revealing its decision-making processes and shedding light on interactions among different agents. To facilitate quick identification of critical states, we have devised multiple visualization views, complemented by a traffic simulation module that allows users to replay specific training scenarios. To validate the utility of our proposed system, we present three comprehensive case studies, incorporate insights from domain experts through interviews, and conduct a user study. These collective efforts underscore the feasibility and effectiveness of {\name} in enhancing our understanding of MARL-based TSC systems and pave the way for more informed and efficient traffic management strategies.}

\end{abstract}

}

\maketitle

\IEEEdisplaynontitleabstractindextext

%
\IEEEpeerreviewmaketitle


\ifCLASSOPTIONcompsoc
\IEEEraisesectionheading{\section{Introduction}\label{sec:introduction}}
\else
\section{Introduction}
\label{sec:introduction}
\fi
\IEEEPARstart{T}{raffic} congestion is a critical and challenging issue in modern urban development, which leads to significant financial losses and detrimental emissions~\cite{deng2023survey,airpollution}. The implementation of intelligent traffic signal control (TSC) presents a practical approach to mitigating traffic congestion. This strategy is a fundamental element of smart cities and has garnered considerable attention from both industrial and government sectors globally, with an aim to identify efficient TSC methods.

\par Despite the widespread adoption of certain adaptive TSC systems, their effectiveness remains heavily reliant on traffic flow models and expert knowledge. This reliance poses limitations on their capacity to dynamically adjust to real-time traffic conditions~\cite{tan2019cooperative,mao2022comparison}. In recent times, reinforcement learning (RL) has emerged as a promising approach to resolving TSC issues, owing to its ability to learn optimal TSC strategies through trial-and-error without being constrained by pre-set rules~\cite{wei2021recent}. While prior research has dedicated efforts to evaluating the performance of RL-based TSC techniques for managing traffic at multiple intersections~\cite{wei2018intellilight,wei2019presslight,xu2021hierarchically}, there has been a notable lack of focus on the decision-making process employed by these models. Additionally, previous studies have primarily relied on a limited set of metrics, such as reward, queue length, and average travel time, to assess the efficacy of TSC models. This approach often lacks the depth required to comprehend the intricate learning mechanisms inherent to RL-based methods. It is essential to note that in real-world scenarios, the trial-and-error nature of TSC models can potentially lead to severe, and in some cases, fatal consequences. Therefore, the development of more interpretable RL models and a comprehensive understanding of the decision-making processes within TSC models are imperative steps towards enhancing their reliability and safety.

\par The field of interpretability in machine learning (ML) is rapidly evolving within the broader scope of artificial intelligence. Visual analytics has emerged as an invaluable tool to enhance the interpretability of ML models~\cite{yuan2021survey}. While there have been extensive studies to unravel the workings of various deep learning models such as Convolutional Neural Networks (CNNs)~\cite{liu2016towards,wang2020cnn,gou2020vatld}, Recurrent Neural Networks (RNNs)~\cite{ming2017understanding,strobelt2017lstmvis} and Generative Adversarial Networks (GANs)~\cite{bau2018gan,kahng2018gan,wang2018ganviz} using visualization techniques, there have been relatively few attempts to apply these methods to enhance the interpretability of RL. \revisedOld{This gap in research primarily stems from the limited exploration of RL models and scenarios~\cite{wells2021explainable}. Researchers have thus far predominantly employed video games~\cite{wang2018dqnviz,wang2021visual,metz2023visitor}, robot control~\cite{he2020dynamicsexplorer,luo2022visualizing,cheng2022acmviz}, and simple 2D environments~\cite{saldanha2019relvis,mishra2022not} as their primary domains for visualizing RL models. However, one notable omission in this body of work has been the domain of TSC based on RL, which has not received the attention it deserves. Additionally, prior research has primarily leveraged single-agent RL models as the backend for visualization, whereas TSC operates on multi-agent reinforcement learning (MARL) systems. The utilization of multi-agent systems entails more than just an increase in the number of agents; it also necessitates addressing the intricate interactions that can occur between these agents. These complex interactions present significant challenges and can profoundly impact the overall performance of the system. While some visual analytics systems have been proposed to explain MARL~\cite{kravaris2023explaining,shi2023maddpgviz}, these solutions may not be directly applicable to TSC settings. To the best of our knowledge, no visual analytics solutions tailored explicitly to the needs of both \textbf{RL} and \textbf{TSC researchers} have delved into the intricacies of comparing multiple RL agents within the context of TSC. Closing this research gap requires the development of advanced visualization tools capable of shedding light on the inner workings of RL models and their applications in TSC, especially in multi-agent scenarios.}

\par Nonetheless, three significant challenges must be addressed to gain a comprehensive understanding of RL-based TSC methods. \revisedOld{
\textbf{(1) Providing a Complete Assessment of RL Models.} In transportation research, RL model performance is typically assessed from a holistic road network perspective~\cite{wei2021recent}, often employing commonly used metrics like reward and queue length. Yet, these metrics do not reveal the underlying rationale behind each agent's actions within a given state. This underscores the pressing need for a complete assessment of RL models. \textbf{(2) Unveiling Dynamic Relationships in Multi-Agent TSC.} 
In a multi-agent setting, a single agent's actions can significantly impact the performance of other agents~\cite{bazzan2009opportunities}, consequently influencing the overall model performance. The dynamic interactions among multiple agents make it challenging to examine the intricate relationships among them, including their collaboration and contribution to the model's performance. \textbf{(3) Interpreting Complex Decision-making Process of RL-based TSC.} An RL model comprises various components, including agents, states, and actions. Each agent engages with the environment, learning from its experiences. However, the behaviors of these agents can lead to unpredictable collective patterns that have a substantial impact on the overall model performance. As a result, articulating and interpreting the decision-making process of the model at various levels becomes challenging.}

\par To tackle the aforementioned challenges, we propose a visual analytics system designed to enhance the exploration and interpretability of RL-based TSC models. Our approach initiates with a comprehensive requirement analysis, identifying the essential needs and concerns of domain experts.\revisedOld{We then carefully select a set of metrics, drawing from both prior research and specific requirements, to extract detailed insights that characterize RL models comprehensively. Subsequently, we apply Shapley values to pinpoint important features for different agents and train interpretable models that unveil decision-making rules governing the relationships between states and the actions of each agent. Additionally, we perform a multi-level analysis (episode level, time-period level and time-step level) to gain a deeper understanding of the decision-making process in RL-based TSC. To validate our approach, we conduct three distinct case studies and expert interviews. In summary, the key contributions of our work can be succinctly outlined as follows:}
\begin{compactitem}
    \item \revisedOld{We begin by examining the requirements of domain experts in the context of TSC  and identify key factors while also delving into the interpretability of RL-based TSC models.}
    \item \revisedOld{We introduce a comprehensive visual analysis system designed to facilitate an in-depth exploration of MARL models within TSC scenarios. This system allows users to evaluate MARL models from a variety of angles, enhancing their understanding.}
    \item \revisedOld{We employ a combination of methods including case studies, expert interviews, and a user study to assess the effectiveness and user-friendliness of our system, thereby providing a well-rounded evaluation of our approach.}
\end{compactitem}

%
%
%
%


\section{Related Work}
\par This section presents two relevant topics, \textit{Reinforcement Learning for Traffic Signal Control} and \textit{Interpretability of Reinforcement Learning}.

\subsection{Reinforcement Learning for Traffic Signal Control}
\par \revisedOld{The traditional pre-set signal control scheme struggles to efficiently handle complex traffic situations. As a result, researchers have proposed adaptive signal control methods that can dynamically adjust signal timing to alleviate traffic congestion~\cite{hunt1981scoot,lowrie1990scats}. RL has emerged as one of the most promising techniques for TSC. It allows control strategies to adapt in real-time based on the current traffic conditions, making it a valuable approach~\cite{wei2021recent}.}

\par In early time, the application of RL-based TSC often involved simplifying the modeling process by focusing on a single intersection. For instance, Shashi et al.~\cite{shashi2021study} employed the Deep Q-Network (DQN) algorithm to adjust traffic signals, making decisions based on LSTM-generated suboptimal actions. Wei et al.~\cite{wei2018intellilight} enhanced the DQN model by incorporating a phase-gated learning model, providing better insights into how the model adapts to real traffic conditions. Many studies have reported favorable results when applying RL control to individual intersections~\cite{9646417,9806135,wei2019presslight}. However, extending this approach to address multi-intersection challenges presents difficulties. It necessitates the coordination of control models for each intersection and involves coping with the high dimensionality resulting from multiple data sources. To tackle this, Wang et al.~\cite{9240060} employed a directional adjacency graph to model the collaborative mechanism among multiple signal lights. Xu et al.~\cite{wei2019colight} introduced graph attention networks for the first time in TSC scenarios, enhancing information exchange among agents. 


\par While prior studies have significantly advanced the field of TSC, a common shortcoming is the insufficient explanation of their models. Bridging the gap between these advancements and practical applications requires researchers to provide compelling justifications for their methodologies. Hence, the primary goal of this study is to not only contribute to the progress in TSC but also to address the need for enhanced user understanding of the model's decision-making process.

\subsection{Interpretability of Reinforcement Learning}
\par In recent years, there has been a growing interest in enhancing the interpretability of artificial intelligence (AI) research. Most of this research has been concentrated on improving the interpretability of ML models, particularly in the domains of classification, decision-making, and action selection, as documented in studies such as~\cite{wells2021explainable,8466590,9233366}. However, compared to ML, achieving interpretability in the context of RL poses a more intricate challenge. RL relies on feedback mechanisms to guide agents in learning optimal actions within their environment, rather than solely relying on predefined training data.

\par Significant progress has been made in enhancing the interpretability of RL models. However, there exists a noticeable gap in research focusing on RL-based TSC scenarios. To address this, we introduced a taxonomy of related work inspired by Stephanie et al.~\cite{milani2022survey}, outlining three key approaches to boost the interpretability of RL models: \textbf{1) Analyzing Feature Importance:} This approach strives to identify the features influencing agents' actions for a given state. Hayes et al.~\cite{hayes2017improving} directly generated explanations by posing questions like ``when do you do?'' to gain insights into the policies employed by RL agents. \revisedOld{Rizzo et al.~\cite{rizzo2019reinforcement} pioneered the use of SHapley Additive exPlanations (SHAP)~\cite{lundberg2017unified} in RL-based TSC, examining the relationship between traffic conditions and agent behaviors.} \textbf{2) Analyzing the Learning Process and Markov Decision Process (MDP):} This approach breaks down rewards, decomposing them to gain insights into an agent's action preferences. For instance, Zhang et al.~\cite{wang2020shapley} proposed the Shapley Q-value algorithm, which dissects the global reward into local components, facilitating optimal decision-making in a multi-agent context. \textbf{3) Analyzing the Policy Level:} This approach illustrates the long-term behavior of the agent. For instance, Topin et al.~\cite{topin2019generation} introduced Abstract Policy Graphs to summarize policies and explain agents' decisions. Guo et al.~\cite{guo2022explainable} focused on generating action advice, using explanations to enhance the transfer of suboptimal strategies during learning in multi-agent scenarios. \revisedOld{Wollenstein-Betech et al.~\cite{wollenstein2020explainability} utilized Knowledge Compilation and the d-DNNF language to generate the decision process of the traffic light controller. Schreiber et al.~\cite{schreiber2021towards} also applied SHAP to explain how features of the road network influence the agent's action in a given state. However, it's important to note that these studies still face some limitations. First, they do not support multi-level analysis, which could not provide more comprehensive interpretability solutions.} \revisedOld{Moreover, there is a need to provide more effective visual techniques and interaction functions to strengthen the scalability and level of comprehension~\cite{wells2021explainable}.}



\par \revisedOld{To enhance the interpretability of RL models, an increasing number of visualization techniques are being employed. For instance, Wang et al.~\cite{wang2021visual} used an interactive saliency map to elucidate the learned strategies of agents in Atari games. Likewise, Gou et al.~\cite{wang2018dqnviz} conducted a visual analysis of the DQN algorithm, providing detailed insights into DQN models at four different levels to aid comprehension. They also proposed a visual design for representing time series data generated by the model. Moreover, He et al.~\cite{he2020dynamicsexplorer}, Jaunet et al.~\cite{jaunet2020drlviz}, and Wang et al.~\cite{wang2021visual} introduced systems designed to assist users in exploring, interpreting, and diagnosing RL models based on RNNs. Additionally, Mishra et al.~\cite{mishra2022not} developed PolicyExplainer, a tool aimed at answering common RL policy-related questions. However, these research predominantly focuses on visually interpreting single-agent behavior in 2D games, which is not adequate for analyzing MARL models in complex traffic scenarios. }

\par \revisedOld{Recently, there has been an emerging interest in exploring the interpretability of MARL through visual analytics. For instance, Kravaris et al.~\cite{kravaris2023explaining} introduced a visual component designed to facilitate the exploration of MARL-based air traffic flow management. Additionally, Shi et al.~\cite{shi2023maddpgviz} presented a visual analytics system aimed at analyzing the training process of the MARL model in a grounded communication environment within a 2D world. However, these studies cannot be directly applied to our specific target scenario. Primarily, these works concentrate on demonstrating data collected directly during the model training phase and lack the in-depth analysis needed to unveil the decision-making process of the model. Furthermore, the training environment in the prior systems is fundamentally distinct from that of TSC. In TSC scenarios, researchers frequently resort to utilizing synthetic road networks~\cite{noaeen2022reinforcement}. This necessitates the provision for integrating a customizable traffic simulation module into the backend of the visual analytics system.}

\par This study focuses on gaining insights into MARL in the context of TSC. Our approach involves introducing a visual analytics approach designed to clarify the model's behavior across different levels. This spans explanations at the time-step level, extending to the episode level. The approach involves generating visual summaries for individual agents at specific episodes, illustrating relationships among multiple agents, and presenting the decision-making process that links actions and states.

\section{Preliminaries}
\par In this section, we clarify how we employ RL to model the TSC problem and introduce the RL algorithm tailored to our particular scenario.

\subsection{RL-based TSC Scenario}
\par For an RL-based TSC scenario, each intersection of a road network functions as an agent. Further details, along with specific specifications and parameters, are presented below:

\par \textbf{Definition of Episodes.} \revisedOld{Episodes are composed of sequential time steps that detail the interaction between the agent and the environment. During each time step, the agent receives observations of the current state of the environment, takes a corresponding action, and consequently transitions the environment to a new state. The agent is then rewarded based on its executed action.}

\par \textbf{Definition of States.} In the context of the TSC problem, there have been various proposals for defining the environment state, encompassing factors like waiting time, queue length, and vehicle location. However, recent research, as exemplified in studies such as~\cite{wei2019presslight,zheng2019diagnosing}, has suggested that the incorporation of complex state definitions may not necessarily yield substantial improvements in performance. Instead, it is advisable to employ simpler state definitions, such as queue length, to effectively capture the environmental conditions in the TSC problem. Therefore, in this study, we have chosen to define the state using queue length.

\par \textbf{Reward Function.} While the fundamental goal of TSC is to reduce the average travel time for all vehicles, using travel time directly as a metric in the reward function is not deemed suitable. This is because travel time doesn't immediately reflect the impact of TSC, and vehicle movements can influence their travel time in complex ways~\cite{wei2021recent}. \revisedOld{Instead, a widely adopted factor in the reward function by TSC researchers is queue length~\cite{wei2019survey}. To address this concern, we have incorporated the queue length of the entire road network as a component of the reward function. Larger queue lengths signify more congested road networks, resulting in lower rewards. Furthermore, from a safety standpoint, it is preferable to minimize the frequency of phase changes by traffic lights and extend the duration of the current phase. Therefore, we have also taken into account the frequency of phase changes when designing the reward function. The reward function for the $i^{th}$ intersection (agent) is formulated as follows:}
\begin{equation}
  \revisedOld{Reward_i = -(\omega_{queue}Queue_i + \omega_{phase}\delta_i)}
\end{equation}
\par In the reward function, $Queue_i$ represents the queue length of the $i^{th}$ agent, which is the cumulative queue length of its entries. Additionally, $\delta_i$ is a binary variable used to penalize phase changes. When an agent switches its current phase, $\delta_i$ is set to 1; otherwise, it remains 0. The frequency of phase changes by an agent determines the extent of penalty it incurs over a given time period. The hyperparameters $\omega_{queue}$ and $\omega_{phase}$ are introduced, with values set at $\omega_{queue}=1$ and $\omega_{phase}=2$ in this study, based on empirical findings.

\par \textbf{Definition of Actions.} In TSC, researchers commonly adopt one of two action settings. The first approach involves maintaining or altering the current phase, which defines the allowed traffic movements at the intersection, within a pre-defined and fixed phase sequence~\cite{mannion2016experimental}. Conversely, the second approach entails selecting a specific phase from a pre-defined yet variable phase set~\cite{arel2010reinforcement}. Given that the first action setting closely mirrors real-world scenarios, we have selected this approach to demonstrate the efficacy of our model. In our designed scenario, the phase sequence consists of two types of phases: ``allow North-South (N-S) to pass'' and ``allow West-East (W-E) to pass''. Furthermore, to minimize frequent and abrupt traffic light changes, each green light phase has been set to a duration of 10 seconds. As previously mentioned, to mitigate potential hazards arising from uncertain and frequent traffic light transitions, whenever an agent switches its current phase to a different one, the intersection's phase remains fully red for the initial three seconds of the 10-second interval, allowing the intersection to clear. In summary, the action is limited to a binary value of 0 or 1, corresponding to the index of a phase in the predefined phase set.




\subsection{Multi-Agent Deep Deterministic Policy Gradient}
\label{sec:MADDPG}
\par \revised{The Multi-Agent Deep Deterministic Policy Gradient (MADDPG)~\cite{lowe2017multi} is a MARL algorithm developed by OpenAI\footnote{\url{https://openai.com/}}. It serves as an extension of the Deep Deterministic Policy Gradient (DDPG)~\cite{lillicrap2015continuous}, specifically designed to address scenarios involving multiple agents. Unlike alternative approaches such as Independent Q-learning (IQL)~\cite{tampuu2017multiagent} and QMIX~\cite{rashid2020monotonic}, MADDPG employs the Centralized Training Decentralized Execution (CTDE) strategy~\cite{lowe2017multi}, a widely used paradigm for large-scale multi-agent training. This approach leverages the Advantage Actor-Critic (A2C) framework~\cite{Paczolay2020New}. Each agent has two independent neural networks called actor and critic to make decisions. Specifically, each agent utilizes the actor to observe local states and take action. During the training process, the critic makes use of global information, such as states and actions of other agents, to evaluate the agent's current action, which can guide model training effectively. MADDPG has been widely used in many fields. Regarding TSC, {\cite{chu2019multi}} has proved that MADDPG performs better against other traditional RL-based TSC algorithms. Consequently, we have chosen MADDPG as the primary algorithm for our study.}

\section{Observational Study}
\subsection{Experts' Current Practices and Bottlenecks}
\label{section:expertPractices}
\par \revisedOld{To gain practical insights into the TSC problem, specifically the objective of minimizing average travel time, we conducted interviews with a panel of six domain experts (E1-E6). These experts encompassed a diverse range of expertise, including two professors with over a decade of experience in TSC and RL (E1-E2), three researchers who had been actively incorporating RL models into their daily work for more than two years (E3-E5), and an expert (E6) specializing in multi-agent control and communication for approximately 10 years. The valuable feedback gathered from these experts played a pivotal role in helping us grasp the existing limitations of RL-based TSC. Throughout the task analysis phase, we sought their guidance on various aspects, including setting up the TSC scenario and identifying key priorities in training the RL model.}

\par \revisedOld{\textbf{Challenges in Real-World Traffic Control and Simulation Tools.} During our interviews with experts, two significant challenges were emphasized. The first challenge pertains to the complexity of the traffic environment, which makes it difficult to obtain all the necessary data from real-world road networks without advanced roadside equipment. The second challenge relates to concerns regarding traffic safety and order, which make it impractical to test algorithms in real-world scenarios. Consequently, researchers have turned to simulation platforms such as \textit{CityFlow}\footnote{\url{https://cityflow-project.github.io/}} and \textit{SUMO}\footnote{\url{https://www.eclipse.org/sumo/}} as invaluable tools for conducting experiments and evaluations in a controlled environment.}

\par \revisedOld{\textbf{Current Workflow for Researchers in TSC.} The experts we consulted view RL as a promising technique for tackling TSC challenges. However, there remains significant potential for improving the model's exploration, evaluation, and interpretability. The current workflow for TSC researchers can be summarized into three key aspects:}

\par \revisedOld{\textit{(1) Setting up a Simulation Environment.} Given the inherent complexity and variability of real-world traffic issues, researchers often choose to abstract the research problems and simplify the road networks. A comprehensive review by Noaeen et al.~\cite{noaeen2022reinforcement} highlighted that around 70\% of the examined TSC studies incorporated synthetic road networks.}

\par \revisedOld{\textit{(2) Model Training.} During the training of RL models, researchers monitor pre-selected critical metrics such as travel time and reward to assess model performance. Typically, they generate basic line charts or output training results directly to the terminal to visualize how these metrics evolve over episodes. As stated by E1, ``\textit{we attempted to use TensorBoard\footnote{\url{https://www.tensorflow.org/tensorboard}} to monitor the training process to capture the connection between action and state, but this approach falls short in providing a comprehensive understanding of the model.}'' Researchers are generally in search of more robust ways to explore the training process and offer detailed insights for each episode.}

\par \revisedOld{\textit{(3) Model Evaluation.} Similarly, researchers have primarily focused on enhancing the algorithm's performance while giving limited attention to elucidating the decision-making process of the model. They have also made efforts to analyze specific states by inspecting agent actions and metrics, particularly rewards and $Q$-values, in these states. The approaches mentioned above suffer from two main drawbacks. First, they lack convenience in identifying agents and states of interest. While traffic simulation modules provided by platforms like \textit{CityFlow} and \textit{SUMO} allow researchers to directly observe states and agents, they do not provide a comprehensive overview of agent policies or enable the rapid identification of specific states. Second, these methods fall short in delivering an efficient and systematic means of illustrating the model's decision-making process and the interactions among agents.}


\begin{figure*}[h]
  \centering
  \includegraphics[width=\textwidth]{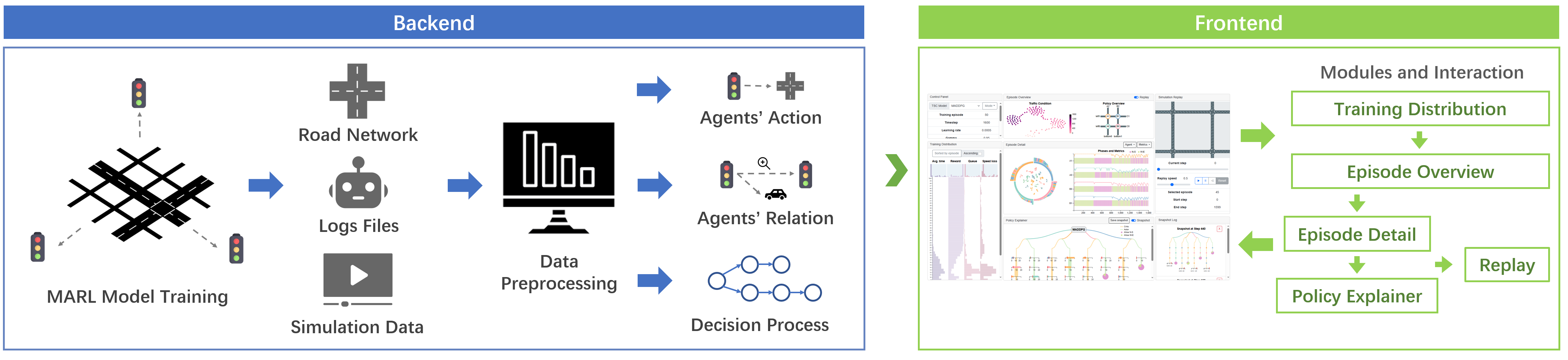}
  \vspace{-6mm}
  \caption{The system pipeline of {\name}. In the back-end engine, we extract critical information about the agents’ behavior, relationships, and decision-making processes from three distinct types of data collected. In the front-end visualization, we offer five coordinated views with rich interactions to facilitate exploration and comprehension of the MARL model.}
  \label{fig:pipeline}
    \vspace{-3mm}
\end{figure*}

\subsection{Experts' Needs and Expectations}
\par \revisedOld{Based on the discussion above, we summarize the following four design requirements, each pertaining to different granularities:}

\par \revisedOld{\prefix{\texttt{\textbf{[R1]}}}\textbf{Episode level: Summarizing the RL Model's Training Process.} Monitoring the training process of the RL model is a critical aspect for researchers in assessing its performance. This monitoring typically involves the observation of key metrics like reward and travel time, which assist in refining the model by detecting changes in these metrics. Thus, it is imperative to offer a comprehensive summary of the model's training process to facilitate a thorough evaluation of its performance.}

\par \revisedOld{\prefix{\texttt{\textbf{[R2]}}}\textbf{Time-period level: Offering a Policy Overview for Each Agent within a Specified Time Frame.} \prefix{\texttt{\textbf{[R2.1]}}}\textit{Describing Temporal Variations in Traffic Conditions in the Road Network:} Across each episode, the traffic conditions in the road network exhibit temporal changes. Given the numerous time steps within an episode, it becomes cumbersome to inspect all these steps individually to analyze the evolving traffic situation. Hence, it becomes essential to provide an overview that aids researchers in comprehending the fluctuations in traffic conditions. \prefix{\texttt{\textbf{[R2.2]}}}\textit{Providing an Overview of Each Agent's Policy:} Across different time periods, the traffic conditions within the road network vary, consequently leading to changes in the policies adopted by the agents. As articulated by E2, ``\textit{It would be useful if an agent's policy can be directly displayed.}'' Additionally, the policy of each agent effectively reflects the model's training progress. For these reasons, our approach should summarize and present each agent's policy within a given time frame.}


\par \revisedOld{\prefix{\texttt{\textbf{[R3]}}}\textbf{Time-step level: Investigating States, Actions, Metrics, and Agent Interactions at Each Time Step.} \prefix{\texttt{\textbf{[R3.1]}}}\textit{Providing a Visual Summary for Each Agent:} Amidst the training of RL models, a substantial volume of data is generated, encompassing actions, states, and rewards for each agent. This abundance of data can pose challenges in comprehending the intricacies of the model. Researchers find value in obtaining a comprehensive overview of an episode, enabling them to swiftly identify anomalous states. Additionally, having access to general information about each agent proves advantageous. Therefore, there's a requirement to furnish a visual summary for each agent, facilitating efficient analysis of the training data. \prefix{\texttt{\textbf{[R3.2]}}}\textit{Demonstrating the Influence and Relationships Among Multiple Agents:} In MARL, a CTDE framework is frequently employed for model training. Within this framework, a centralized controller guides each agent based on global state, action, and reward information. Decisions made by one agent can influence others, and their actions may have repercussions on fellow agents. Consequently, as suggested by E3 and E4, it becomes crucial to consider the interactions among different agents and analyze their impact on the model's overall performance. Thus, there's a necessity to demonstrate the influence and relationships among multiple agents, contributing to a deeper comprehension of the model's behavior. \prefix{\texttt{\textbf{[R3.3]}}}\textit{Presenting the Consistent Rules Governing Actions and States:} Training a MARL model can be a complex undertaking, sometimes resulting in agents exhibiting unpredictable behavior without clear rationales for their actions. As stated by E5, ``\textit{Sometimes agents become uncontrollable and I don't know what happened exactly.}'' In response to this challenge, the experts recommend that our system should assist in exploring the decision-making process of RL models, particularly by facilitating a better understanding of the connection between an agent's actions and the underlying state. Hence, it becomes imperative to present the consistent rules governing actions and states, aiding in the comprehension of agent behavior.}

\par \revisedOld{\prefix{\texttt{\textbf{[R4]}}}\textbf{Enabling Simulation Progress Replay.} Understanding the decision-making process of an RL model based solely on data can be a formidable challenge. In the realm of TSC, researchers frequently rely on simulation platforms for training RL models, as this approach simplifies the replication of specific scenarios post-training. E2 and E5 emphasized the significance of observing agent behavior in specific states, underscoring the need for a replay module to aid researchers in scrutinizing agent behaviors within particular contexts. Consequently, it becomes imperative to grant users the capability to replay the simulation progress, facilitating a more profound comprehension of the model's decision-making process.}

\section{Overview of {\name}}
\par In the system pipeline, depicted in~\cref{fig:pipeline}, we have two primary components: the \textit{back-end engine} and the \textit{front-end visualization}. Within the back-end engine, our initial steps involve establishing a TSC scenario and training a MARL model. This entails inputting the road network, traffic flow, and optionally, a set of phases, followed by fine-tuning the hyperparameters until the selected model effectively controls the traffic lights and attains a satisfactory performance level. Subsequently, we extract the outputs from the trained networks. This extracted data then undergoes preprocessing to enable analysis of various aspects, including the agents' actions, their interactions, and the decision-making process. On the front-end visualization side, we have meticulously crafted five coordinated views equipped with rich interactive features, which allow users to comprehend the model's decision-making process more effectively.

\section{Back-end Engine of {\name}}
\par In this section, we will elucidate the data processing procedures, encompassing scene initialization, feature extraction, and the subsequent outputs derived from these processes.



\subsection{Scene Initialization}
\label{sec:scene}
\par Using the \textit{Libsignal}\footnote{\url{https://darl-libsignal.github.io/}}~\cite{HaoMei2022LibSignalAO} platform as a basis, we establish a scenario comprising a 2$\times$2 network, where each intersection serves as an agent denoted by A0, A1, B0, and B1 (\cref{fig:teaser}$\rm C_2$). \revisedOld{During our expert interviews, researchers typically overlook specific details like turning vehicles when studying general TSC problems of road networks. Consequently, our traffic flows are restricted to straight movement only.} Moreover, we assign unique identification numbers to each node. An edge is represented by two nodes. For instance, ``A0A1'' refers to the edge connecting the intersection from node A0 to A1.

\par After approximately $50$ training episodes using MADDPG with specific training hyperparameters, such as a learning rate of 0.0005 and a target network update frequency of once every 10 episodes, the algorithm demonstrates a tendency toward convergence. Each training episode is segmented into $1600$ time steps, with each step corresponding to one second of simulation time. \revisedOld{Additionally, we divide the traffic flow into four distinct stages to observe agent behavior under varying traffic conditions. Each stage has a duration of $400$ time steps. The four stages, outlined in \cref{tab:traffic-flow-settings}, are: \textbf{(1) W-E Only}, \textbf{(2) N-S Only}, \textbf{(3) N-S lower than W-E}, and \textbf{(4) W-E lower than N-S}. The traffic flow in the first two stages moves in only one direction, while the last two stages involve traffic flow in all directions, with a larger flow in one specific direction.}


\begin{table}
    \centering
    \caption{\revisedOld{The traffic flow settings of our cases.}}
    \label{tab:traffic-flow-settings}
    \begin{tabular}{c c c}
        \hline
        \textbf{Stage} & Direction & Traffic Flow (veh/h) \\
        \hline
        \multirow{2}{*}{1} & W-E & 1800 \\
          & N-S & 0 \\

        \hline
        \multirow{2}{*}{2} & W-E & 0 \\
          & N-S & 1800 \\

        \hline
        \multirow{2}{*}{3} & W-E & 1800 \\
          & N-S & 600 \\

        \hline
        \multirow{2}{*}{4} & W-E & 600 \\
          & N-S & 1800 \\
          
        \hline
    \end{tabular}
\end{table}


\par Upon completion of each training episode, the trained model will undergo a testing phase to evaluate whether it has entered a state of over-fitting. The testing scenario will be identical to the training scenario.

\subsection{Data Description}
\label{sec:analyticsTasks}
\par To provide explanations of the MADDPG algorithm, we have extracted data from the training and testing models as well as logger files. The extracted data can be broadly categorized into three groups: \textbf{1) Model Data.} This group includes the outputs of both the training and testing models, which contain various neural network metrics. As described in Section~\ref{sec:MADDPG}, the MADDPG algorithm creates a critic network and an actor network for each agent. We have extracted the input and output data from these networks. Specifically, for the critic network, the network requires global observation $O$ and action probability $P$ from each agent to score the corresponding agent's behavior with Black box neural networks. As for the actor network, the network predicts the probability of each action based on the local observation of the agent. \textbf{2) Logger Files.} During the model training or testing phase, a large number of logger files are generated. These files contain various metrics such as mean reward, queue length, delay, and travel time, for each episode. In addition, for each episode, we also obtain all the observations such as queue length of each lane, current/last action, and current/last phase for each of the $10$ time steps. \textbf{3) Simulation Data.} \textit{LibSignal} can generate traffic simulation data automatically. The phases of intersections and vehicles' location will be recorded every time step. Simulation data is useful to replay the status of the road network for detailed analysis.



\subsection{Feature Extraction}
\par To identify the key features, we perform additional data processing from two different perspectives.
\par \textbf{Decision-making Process.} \revisedOld{In the A2C framework, each agent utilizes neural networks for decision-making. Drawing inspiration from Mishra et al.~\cite{mishra2022not}, we independently train decision trees for both the critic and actor networks to gain a deeper understanding of how neural networks make decisions. Specifically, we employ all global information and action probabilities from all agents to train a regression tree. This regression tree establishes a connection between global information and the agent's evaluation of the current state, providing insights into which features contribute to a specific action. Importantly, this method is model-agnostic, utilizing only state and action information as inputs without introducing new structures inside the model.}
\par \textbf{Feature Importance.} \revisedOld{Another model-agnostic approach we employ to glean insights into the model training process is SHAP~\cite{lundberg2017unified}, commonly used for measuring feature importance. Previous studies~\cite{rizzo2019reinforcement} and~\cite{schreiber2021towards} have applied SHAP to RL-based TSC. Building on their work, we also use SHAP to analyze the neural networks of each agent and illustrate feature importance.}

\begin{figure*}[h]
  \centering
  \includegraphics[width=\linewidth]{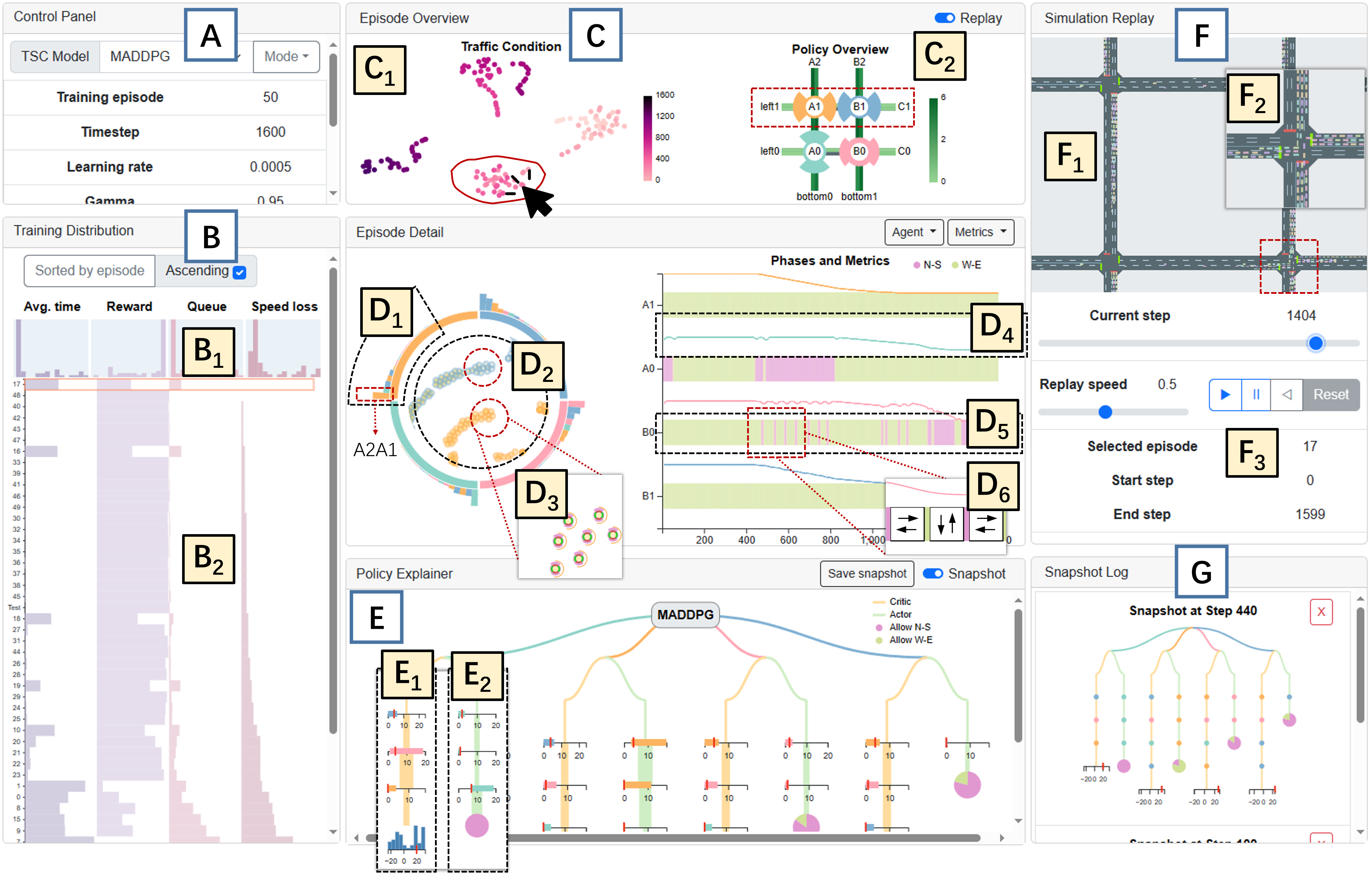}
  \vspace{-6mm}
  \caption{\revised{Our visualization system, {\name}, provides an in-depth analysis of MARL models in TSC scenarios.} The {\controlpanelCapital} (A) presents parameters in model training and model testing. The {\trainingdistributionCapital} (B) provides the distribution of the metrics and ranks the episode based on the metrics. The {\episodeoverviewCapital} \revisedOld{(C) presents a summary of traffic conditions and each agent's policy at a certain episode. The {\episodedetailCapital} (D) provides a visual summary for each agent in an episode, including information of the state, action, and selected metrics, and demonstrates relationships among multiple agents.} The {\policyexplainerCapital} (E) provides explanations between local state and action, global information and critic value. The {\simulationreplayCapital} (F) supports the replay of an arbitrary episode or time step in the simulation situation. The {\snapshotlogCapital} (G) saves the snapshots of the {\policyexplainerCapital}.}
  \label{fig:teaser}
    \vspace{-3mm}
\end{figure*}

\section{Front-end Visualization of {\name}}
\par In this section, we present the visual design of each view within {\name}, accompanied by an exploration of diverse design alternatives that underwent evaluation. Our system comprises six primary components: the {\controlpanelCapital}, {\trainingdistributionCapital}, {\episodeoverviewCapital}, {\episodedetailCapital}, {\policyexplainerCapital} \& {\snapshotlogCapital}, and {\simulationreplayCapital}.

\subsection{\controlpanelCapital}
\par The {\controlpanelCapital} (\cref{fig:teaser}A) is designed to serve as a hub for presenting training information related to various models. Each model's profile encapsulates details regarding both the training and testing processes. This includes essential information such as the training episode, time step, and crucial hyperparameters like the learning rate. Moreover, the {\controlpanelCapital} affords users the flexibility to seamlessly switch between the display of training and testing data, with the system initially set to showcase the training data as the default view. When a user selects a specific RL model of interest, the relevant information dynamically updates below, while other associated views also receive corresponding updates.

\subsection{{\trainingdistributionCapital}}
\par To offer a concise overview of the model training process (\prefix{\texttt{\textbf{[R1]}}}), commonly used metrics are employed to evaluate the traffic condition of the road network. These metrics encompass parameters like \textit{speed loss}, \textit{queue length}, \textit{reward}, and \textit{avg. time}, all of which were exclusively derived from the training data. To analyze the distribution of these metrics effectively, we devise the {\trainingdistributionCapital} (\cref{fig:teaser}B), implementing the lineup design~\cite{gratzl2013lineup}. This design choice enables users to rapidly visualize the distribution of the four metrics and swiftly locate their desired episode within the data. Given the substantial differences in the value ranges of these metrics, we conduct data normalization to ensure meaningful comparisons. Furthermore, users are allowed to arrange and sort specific metrics as per their requirements.

\subsection{\episodeoverviewCapital}
\par To offer a comprehensive policy overview for each agent within a specific time frame (\prefix{\texttt{\textbf{[R2]}}}), the {\episodeoverviewCapital} has been intricately designed to facilitate an in-depth exploration spanning various time stages and policies. This component comprises two main sections: \textit{Traffic Condition} and \textit{Policy Overview}.
\par \textbf{Traffic Condition.} To observe how traffic conditions evolve throughout an episode (\prefix{\texttt{\textbf{[R2.1]}}}), we count the number of vehicles present on each road and employ the \textit{t-SNE} algorithm~\cite{van2008visualizing} to project traffic conditions onto a 2D plane. Each time step's traffic condition is represented as a point in~\cref{fig:teaser}$\rm C_1$. The color of each point corresponds to its respective time step, with lighter colors denoting earlier times and darker colors representing later times. The use of \textit{t-SNE} ensures that similar traffic conditions are visually clustered together.
\par \textbf{Policy Overview.} For a comprehensive understanding of each agent's policy (\prefix{\texttt{\textbf{[R2.2]}}}), we construct a representation of the road network, as depicted in \cref{fig:teaser}$\rm C_2$. In this representation, the opacity of the green color is indicative of the number of vehicles on the road. Each intersection in the diagram represents an agent. Within each intersection, we employ four sectors to illustrate the probability of allowing vehicles to pass, corresponding to the four cardinal directions. The size of each sector correlates with the probability it represents, with larger sectors indicating higher probabilities.
\par Moreover, users can select a group of points of interest with a lasso in \cref{fig:teaser}$\rm C_1$. When points are selected, the corresponding sectors in the four directions, as well as the road network, are updated to provide further insights.

\subsection{\episodedetailCapital}
\par \revisedOld{In order to provide a visual summary for each agent (\prefix{\texttt{\textbf{[R2]}}}) and demonstrate the influence and relationship among multiple agents (\prefix{\texttt{\textbf{[R3]}}}), the {\episodedetailCapital} is designed to provide a thorough exploration of each episode, which mainly contains two parts, namely \textit{state projection and feature importance} and \textit{traffic signal phase and metrics}.}

\par
\begin{figure}[h]
  \centering
  \includegraphics[width=0.5\textwidth]{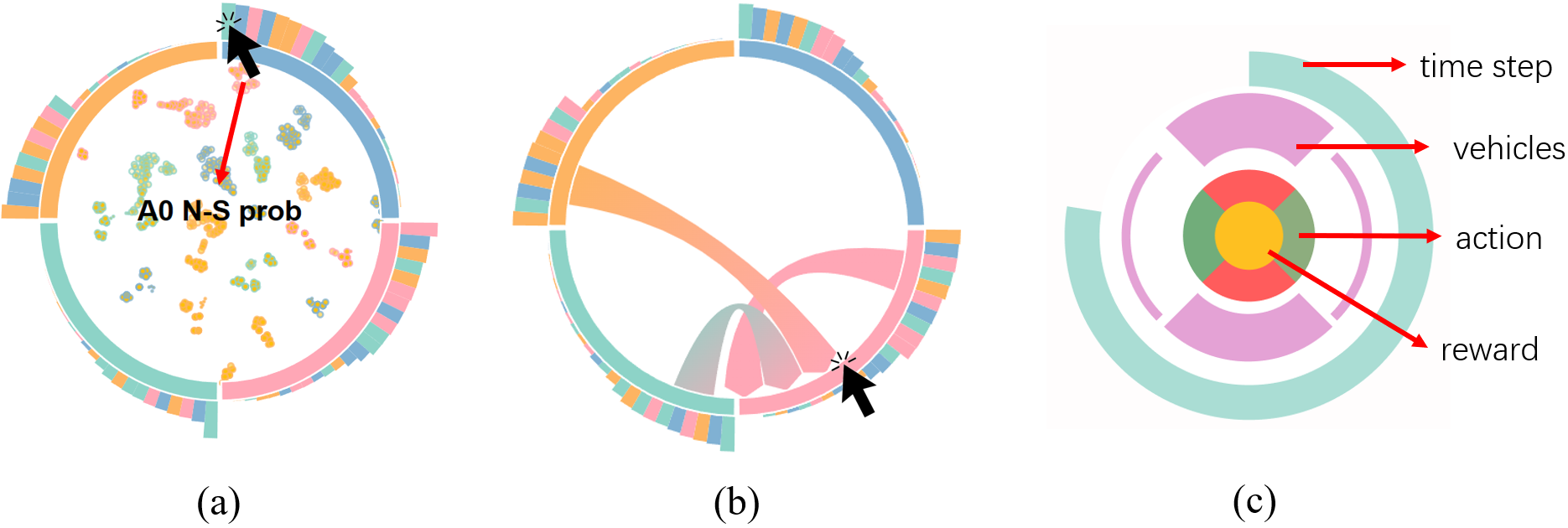}
  \vspace{-6mm}
  \caption{\revisedOld{Glyph design and interaction in the \episodedetailCapital. (a) When hovering the mouse over the bar charts (feature importance), the corresponding feature's name is displayed in the center. (b) Utilization of a chord diagram to illustrate the relationships among agents. (c) The glyph design showcasing state information, employing distinct rings to represent various aspects of the episode, such as time step, traffic flow, action, and reward.}}
  \label{fig:EpisodeDetail-chord-glyph}
\end{figure}

 \textbf{State Projection and Feature Importance.} On the left part, a circular segment design is adopted to present information about the four agents in our scene (\cref{fig:teaser}$\rm D_1$). A ring is partitioned into four segments (arcs) and their colors correspond to different agents. Inside the ring, we utilize the~\textit{t-SNE} algorithm to project the state information, which includes the current decision action, rewards, current time step, and the traffic condition of the intersection on the map, into the 2D plane (\cref{fig:teaser}$\rm D_2$). This approach ensures that similar states are situated closer to each other. Before zooming in, the color of each state point corresponds to the color of the agent. The opacity of the point encodes the reward. After zooming in, to provide more state information, we design a glyph, as shown in~\cref{fig:EpisodeDetail-chord-glyph}c. The center circle encodes reward, with darker shades representing higher rewards. The inside ring encodes action, with red representing stop and green representing pass. The middle ring encodes vehicles in four directions, with the wider arc representing more vehicles. The outer ring encodes the time step, with the long arc representing later time step. Users can click on these state points (glyphs) in the 2D plane, and other views (e.g., the {\policyexplainerCapital}) will be updated correspondingly. Moreover, we use a chord diagram to display an overview of the influence among agents (\cref{fig:EpisodeDetail-chord-glyph}b). A chord from agent A to agent B indicates some features of A affect B's decision-making. The width of the chord encodes the number of features. To avoid visual clutters, when users hover over an agent, the chord diagram will only display influences related to this agent. Gradient colors are used to strengthen agent information. Outside the arcs, bar charts are used to depict the interplay between agents and Shapley values of the features. To enable better comparisons, these bar charts are encoded with the same color as the corresponding agents and arranged in order of Shapely values(clockwise). As shown in~\cref{fig:EpisodeDetail-chord-glyph}a, when users hover over a specific feature, the corresponding feature name will be displayed, facilitating a more comprehensive evaluation of each feature's importance in decision-making.

\par \textbf{Traffic Signal Phases and Metrics.} On the right side, to better visualize and compare metrics and actions among different agents, we devise four sets of lines and bands corresponding to each of the four agents. The line chart effectively illustrates the fluctuations in selected metrics (\cref{fig:teaser}$\rm D_4$). Bands along the time axis in~\cref{fig:teaser}$\rm D_5$ represent traffic signal phases, with different colors signifying phases selected by individual agents. Users can zoom in to closely examine metrics and actions at each time step. To facilitate the quick identification of traffic signal phases, when users zoom in to a certain degree, icons appear to directly display traffic signal phases (\cref{fig:teaser}$\rm D_6$). Users have the flexibility to select a time step within the bands to update other views.

\par
\begin{figure}[h]
  \centering
    \vspace{-3mm}
  \includegraphics[width=0.5\textwidth]{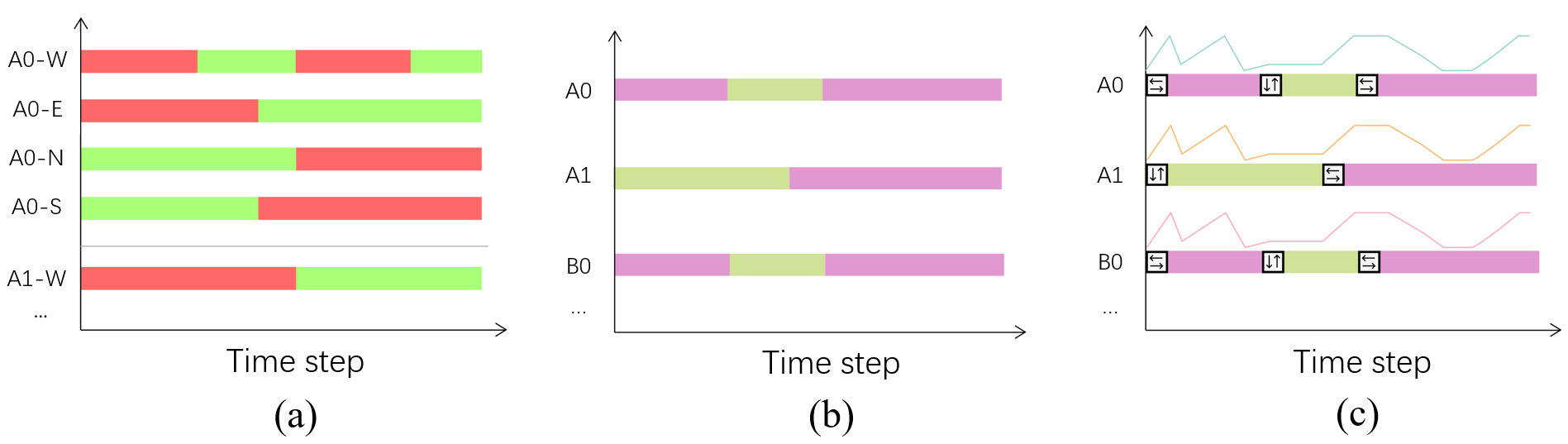}
  \vspace{-6mm}
  \caption{\revised{Various design alternatives were evaluated for the components within the {\episodedetailCapital}.} (a) Represents a commonly employed design in TSC for displaying traffic signal phases. (b) Introduces an alternative design aimed at conserving space and facilitating the comparison of agents' actions. (c) Depicts the current design, employing icons to directly illustrate different traffic signal phases. Line charts are utilized to concurrently compare metrics for different agents.}
  \label{fig:episode-detail-design-alternative}
    \vspace{-3mm}
\end{figure}

 \textbf{Design Alternatives.} Before settling on the current designs, we explored various alternatives. Initially, ~\cref{fig:episode-detail-design-alternative}a was employed to showcase agents' traffic signal phases, which is a common practice in TSC. However, it could only display one traffic signal phase per row, leading to significant space consumption when dealing with numerous agents and traffic signal phases. As illustrated in~\cref{fig:episode-detail-design-alternative}b, an attempt was made to conserve space and facilitate the comparison of actions among different agents by using one row to represent the traffic signal phase information of an agent, with different colors encoding various traffic signal phases. Despite these efforts, this design proved less intuitive as the number of traffic signal phases increased. Subsequently, the final version (\cref{fig:episode-detail-design-alternative}c) was developed to address these challenges. In this version, an icon is placed in front of each bar to enable users to quickly identify different traffic signal phases. The use of line charts effectively illustrates how metrics change over time.

\subsection{{\policyexplainerCapital} \& {\snapshotlogCapital}}
\par In order to facilitate a better understanding of the decision-making process of MARL models, we have adopted a tree-based design to present the policies governing the relationship between actions and states (\prefix{\texttt{\textbf{[R3]}}}). As illustrated in~\cref{fig:teaser}E, the root node of the tree represents a selected ``MADDPG'' model, and four subtrees are used to encode rules for four agents, each designated with its respective agent color. For each agent, a subtree is divided into two branches: 1) The left branch (\cref{fig:teaser}$\rm E_1$) elucidates how a critic arrives at a decision. This involves traversing several feature judgments, represented by feature ranges in the corresponding agent's color, to reach a final critic value encoded with a red bar and distribution. The higher the critic value, the higher the critic's evaluation of the agent's current situation. The current feature value is depicted as a red bar within the feature range. 2) The right branch (\cref{fig:teaser}$\rm E_2$) illustrates the action-performing process of the actor. Similar to the critic branch, it involves navigating through feature judgments before executing an action. The final action distribution is visualized as a pie chart. The width of a path within the tree encodes the value range of the associated features. Additional information is revealed when users hover over feature judgments and the corresponding roads will be highlighted in the road map (\cref{fig:teaser}$\rm C_2$). To support the comparison of different rules, users can save a snapshot to the {\snapshotlogCapital} by clicking the ``Save snapshot'' button. Users can easily reload a snapshot by clicking it. \revised{In addition, users can hide the {\snapshotlogCapital} by clicking the switch button on the top-right of the {\policyexplainerCapital}.}

\par
\begin{figure}[h]
  \centering
  \includegraphics[width=0.5\textwidth]{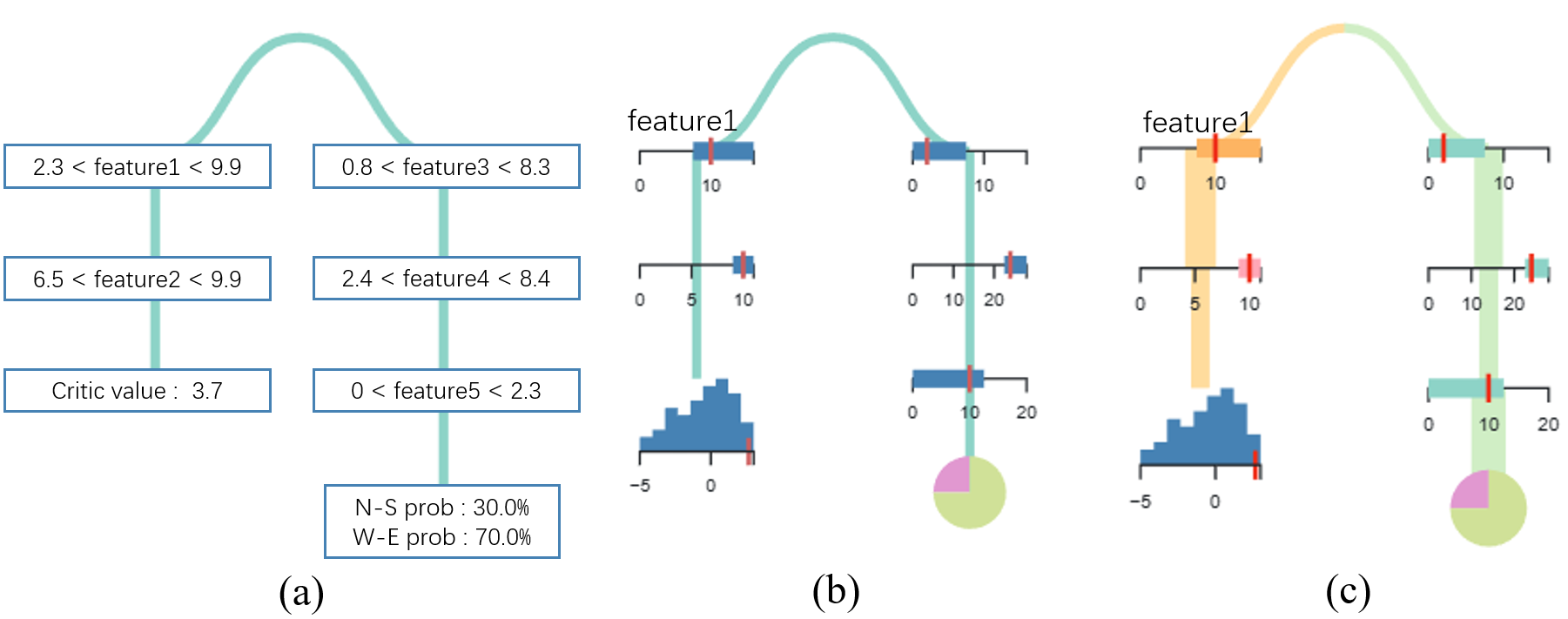}
  \vspace{-6mm}
  \caption{\revised{Design alternatives were considered for the {\policyexplainerCapital} component. (a) One design approach involved presenting feature names and their value ranges directly within each branch, supplemented by accompanying text.} (b) Another design utilized distribution representations but did not differentiate between branches for different agents or vary branch thickness to indicate the number of rules. (c) The current design employs distribution representations with distinct colors for each agent's branches and varying branch thickness to visually represent the number of rules.}
  \label{fig:DA-Explainer}
\end{figure}

 \textbf{Design Alternatives:} Prior to adopting the current tree-based design, several alternative designs were considered. For instance, as shown in~\cref{fig:DA-Explainer}a, one design approach proposed by~\cite{mishra2022not} involved directly displaying feature names and feature value ranges in each branch, accompanied by text explanations. However, this design was found to lack intuitiveness. Another design, depicted in~\cref{fig:DA-Explainer}b, utilized distribution representations but did not adequately distinguish between actor and critic branches, used uniform colors for each agent's branch, and lacked variation in branch thickness. Consequently, the current tree-based design was implemented, introducing distinct colors for agent branches and utilizing branch thickness to visually encode the number of rules, as illustrated in~\cref{fig:DA-Explainer}c.

\subsection{\simulationreplayCapital}
\par To better understand the model's decision-making process (\prefix{\texttt{\textbf{[R4]}}}), we have incorporated the {\simulationreplayCapital} module, making use of the \textit{LibSignal} platform. This module comprises an upper section that displays a road network (\cref{fig:teaser}$\rm F_1$) utilized for simulation purposes. \revisedOld{When users select an episode in the {\trainingdistributionCapital}, the {\simulationreplayCapital} seamlessly transitions to the corresponding episode.} The lower portion of this module (\cref{fig:teaser}$\rm F_3$) allows users to control the simulation. For example, users can pause the replay process at any point and adjust the replay speed within a range from 0.1 to 1, enabling a closer examination of specific moments. Additionally, the {\simulationreplayCapital} supports the option to restart the current process, restoring the original settings for replay purposes. Furthermore, for quick access to a particular time step, users can select a specific state within the {\episodedetailCapital}, which will trigger the display of the simulation process, commencing from the specified time step. \revised{Similarly, users can close the {\simulationreplayCapital} by clicking the switch button on the top-right of the {\episodeoverviewCapital}.}

\begin{figure*}[h]
  \centering
  \includegraphics[width=\textwidth]{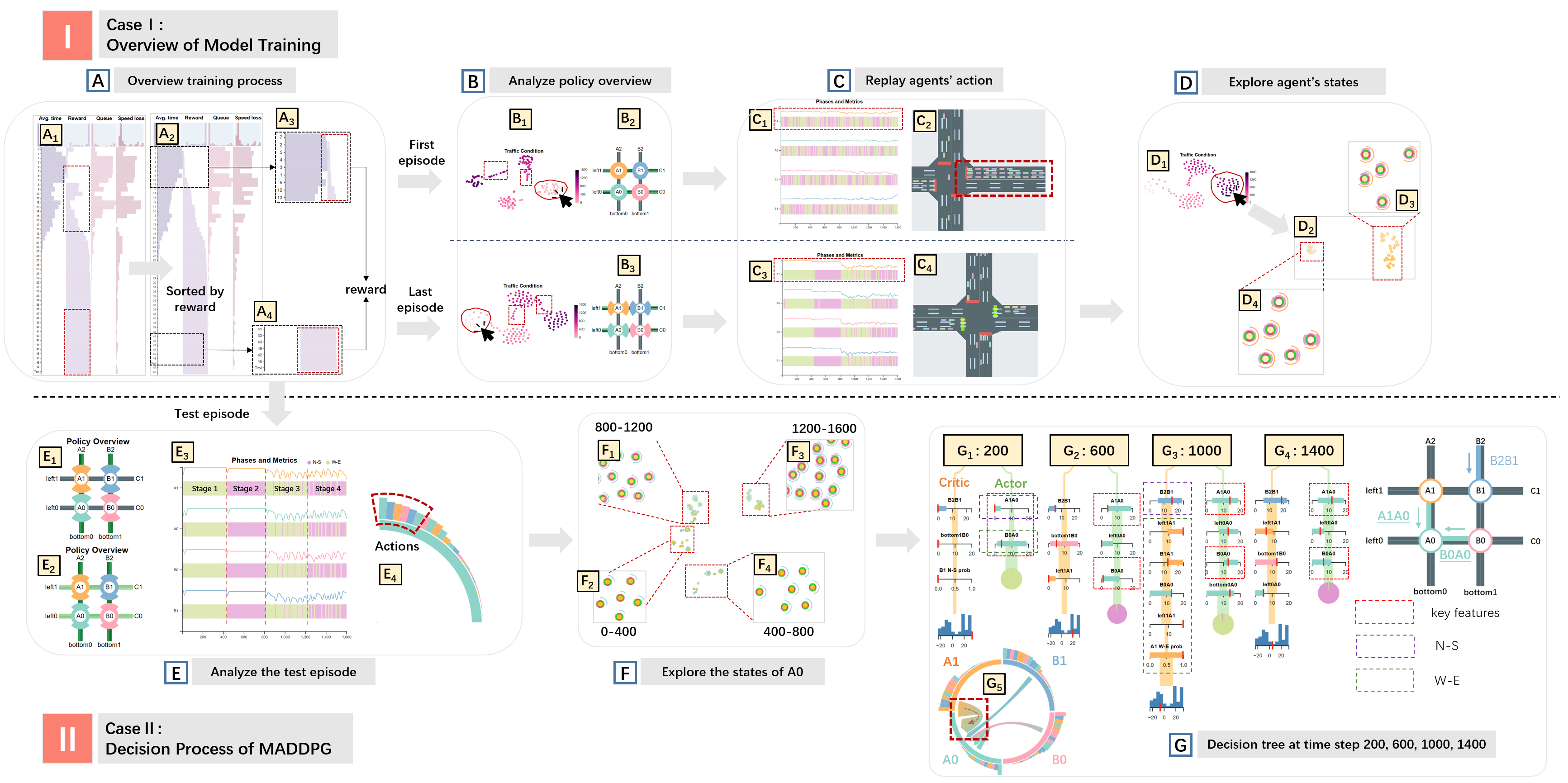}
    \vspace{-6mm}
  \caption{\revised{Experts' operations during the case study I and II. (A) Use the {\trainingdistributionCapital} to compare different episodes. (B) Analyze and compare the overview of agents' policy in different episodes. (C) Replay agents' actions with more details. (D) Select a time range and explore the states of an agent. (E) Analyze how agents react to different traffic flows in the test episode. (F) Explore the states of agent A0. (G) Reveal the decision-making process of agent A0.}}
  \label{fig:case}
  \vspace{-3mm}
\end{figure*}

\section{Evaluation}
\label{section:evaluation}
\par We perform three case studies and conduct interviews with E1-E6, as introduced in~\cref{section:expertPractices}. \revisedOld{\Cref{section:caseStudy} details the cases conducted by E2, E1, and E6 respectively, due to the representativeness of these cases.} \Cref{section:interview} presents the feedback from all experts. Subsequently, we recruit an additional $12$ participants and conduct a user study to assess the performance of {\name} in system workflow, visual design and interaction, as well as system usability, as discussed in~\cref{section:userStudy}.

\subsection{Case Study}
\label{section:caseStudy}

\subsubsection{\textbf{Case I: Overview of Model Training}}
\par \zyt{E2 focuses on crafting RL models and often needs to summarize the performance of model training.  His actions are illustrated in \cref{fig:case}I.}
\par \revised{To gain a macroscopic understanding of the model training situation, he typically examined the \trainingdistributionCapital~(\cref{fig:case}$\rm A_1$). Sequentially analyzing the episodes, he observed that the model training was effective, experiencing some fluctuations in the early stages but becoming more stable later on. Conversely, episodes with larger rewards were predominant in the later stages~(\cref{fig:case}$\rm A_4$), providing further evidence of the model's effectiveness.}

\par Then, E2 navigated to the \episodeoverviewCapital~to scrutinize the agents' policies. Observing the road network's traffic conditions in \cref{fig:case}$\rm B_1$, he identified four distinct clusters corresponding to the four types of traffic flow settings. The red rectangles in~\cref{fig:case}$\rm B_1$ denoted transitions between different traffic flows. E2 chose the first stage of traffic flow (W-E direction only) in the initial episode. In \cref{fig:case}$\rm B_2$, each sector's size was nearly identical, indicating that all agents assigned almost the same probability to the green light in each direction. E2 remarked, ``\textit{Agents appear to take actions randomly, and their strategy for allowing vehicles to pass is confusing and inefficient.}'' In contrast, the last episode in \cref{fig:case}$\rm B_3$ showed larger sectors in the W-E direction than in other directions. All agents permitted only west and east bound vehicles to pass, aligning with the traffic flow's direction.

\par Subsequently, E2 used the {\episodedetailCapital} to assess actions and metrics' temporal changes. He chose agent A1 as an example, focusing on the critic value to observe the model's grasp of global information. For actions, he observed frequent color changes in the bands, signifying continuous policy exploration in the first episode~(\cref{fig:case}$\rm C_1$). Conversely, the strategy stabilized in the last episode~(\cref{fig:case}$\rm C_3$) with A1's actions distinctly segmented into four stages corresponding to the four-stage traffic flow. The line chart in~\cref{fig:case}$\rm C_1$ and $\rm C_2$ revealed that the critic value in the first episode fluctuates within a narrow range, indicating the model's inability to evaluate the entire road network's traffic condition. Conversely, in the last episode, the critic value accurately reflected the road network's traffic conditions. A higher critic value denoted better traffic conditions. E2 leveraged the {\simulationreplayCapital} to directly inspect A1's traffic condition. For instance, in the last traffic flow stage where N-S direction is the primary traffic flow direction, the first episode, due to trial-and-error, sometimes leads to severe congestion~(\cref{fig:case}$\rm C_2$). Conversely, in the last episode, the intersection's traffic condition was more orderly. A1 kept the N-S direction green until queues in the W-E direction accumulated, prioritizing the main traffic flow~(\cref{fig:case}$\rm C_4$).

\par \revisedOld{Continuing with the reward and state analysis, E2 explored A1's states during the last traffic flow stage~(\cref{fig:case}$\rm D_1$). Given the model's proficiency in the last episode, his focus shifted accordingly. Notably, he observed two clusters in \cref{fig:case}$\rm D_2$~, indicating high similarities. The left cluster was smaller than the right, prompting his to zoom in for a more detailed examination. It became evident that the left cluster represented states when A1 activated the green light in the W-E direction. As shown in \cref{fig:case}$\rm D_4$, the larger area of purple sectors in the W-E direction suggested more vehicles waiting in that intersection direction. Conversely, the right cluster denoted states when A1 activated the green light in the N-S direction. The upper part of this cluster indicated lower rewards, signifying congestion in the main traffic flow direction (N-S direction)~(\cref{fig:case}$\rm D_3$).}

\par \revisedOld{In summary, E2 concluded that the model underwent effective training with the increasing number of training rounds. All metrics exhibited substantial improvement, indicating that agents had learned to adapt and employ different strategies in response to varied traffic flow settings.}

\subsubsection{\textbf{Case II:  Decision Process of MADDPG}}
\par \revised{E1 aimed to delve into the decision-making process of the model. Her specific operations are illustrated in~\cref{fig:case}II.}

\par \revised{Considering the decision process is stable in the test episode, E1 conducted a comprehensive analysis. Initially, she observed the metrics of the test episode in the \trainingdistributionCapital~(\cref{fig:case}$\rm A_4$). Despite the test episode not achieving the optimal performance, it remained within the normal fluctuation range.}

\par Then, within the \episodeoverviewCapital, she selected two time periods, as depicted in \cref{fig:case}$\rm E_1$ and $\rm E_2$. These illustrations revealed a correlation between higher traffic flow (darker green on the roads) and a higher probability (larger area of the sectors) of activating green lights in corresponding directions. \revised{\cref{fig:case}$\rm E_3$ provided detailed insights into the agents' actions. For example, in the third stage in {\cref{fig:case}$\rm E_3$}, where the main traffic flow was in the W-E direction (\cref{tab:traffic-flow-settings}), the area of green bands (indicating turning on the green light in the W-E direction) significantly exceeded that of purple bands.} Regarding the inter-agent relationship, E1 observed that higher bars consisted of colors from other agents, indicating that an agent's action was significantly influenced by others (\cref{fig:case}$\rm E_4$). She remarked, ``\textit{Agents tend to observe each other's actions, reflecting the process of cooperation among agents.}''

\par Moving on to the projection of states, E1 examined the projections of each agent separately and found that the results could be roughly divided into four clusters based on locations and rewards. Taking agent A0 as an example, four clusters were highlighted with red rectangles (\cref{fig:case}F). To compare differences within these clusters, she zoomed in on each cluster and observed the glyph in \cref{fig:case}$\rm F_1-F_4$. Comparing the time step and traffic flow of the states in each cluster, she discovered that states within the same cluster were highly similar. \revised{Considering the traffic flow in the middle of a stage is stable without interference, she chose the middle time steps of the four traffic flow stages ($200$, $600$, $1000$, and $1400$) for further exploration.}

\par E1 conducted a detailed analysis of A0's decision process at four representative time steps, utilizing the {\snapshotlogCapital} to save snapshots for easy comparison of agents' decision processes under different states. She identified \textit{A1A0} and \textit{B0A0} as key features influencing the agents' decisions at each traffic flow stage (\cref{fig:case}$\rm G$). Upon comparing the values of these features, she observed that the agent's policy was influenced by the number of vehicles in different directions. For instance, at time step $200$ (\cref{fig:case}$\rm G_1$), the actor branch considered the number of vehicles on \textit{A1A0} and \textit{B0A0}, with the value of \textit{B0A0} significantly higher than \textit{A1A0}. Consequently, the agent chose to allow vehicles in the W-E direction to pass through the intersection. When the number of vehicles in two directions was similar (time step $1000$ in \cref{fig:case}$\rm G_3$), E1 delved deeper into how the critic guided the model's decision process.

\par E1 noted the chord diagram in \cref{fig:case}$\rm G_5$, where the chord from agent A1 to agent A0 was larger than other chords. She remarked, ``\textit{It seems that A1 has a greater impact on A0 at this time.}'' Examining the critic tree in \cref{fig:case}$\rm G_3$, she found an evaluation of the agent's action based on features mainly related to agent A1 in the W-E direction (highlighted with a green dashed rectangle), such as \textit{left1A1} and \textit{A1 W-E prob}. At this time step, A1 turned on the green light in the W-E direction (\textit{A1 W-E prob}) due to a higher number of vehicles in the W-E direction (\textit{left1A1}, \textit{B1A1}). For agent A0, maintaining the N-S direction green might interrupt A1 and lead to congestion in A1's N-S direction. Therefore, agent A0 opted to turn on the green light in the W-E direction.

\par In conclusion, E1 concluded that the agent prioritizes the direction with heavier traffic flow. In cases where traffic situations are similar, the critic serves as a valuable aid in decision-making based on other features. E1 remarked, ``\textit{This looks like priority control in TSC, where traffic lights stay green for the main road until there are too many vehicles waiting in the side road.}''

\begin{figure*}[h]
  \centering
  \includegraphics[width=\textwidth]{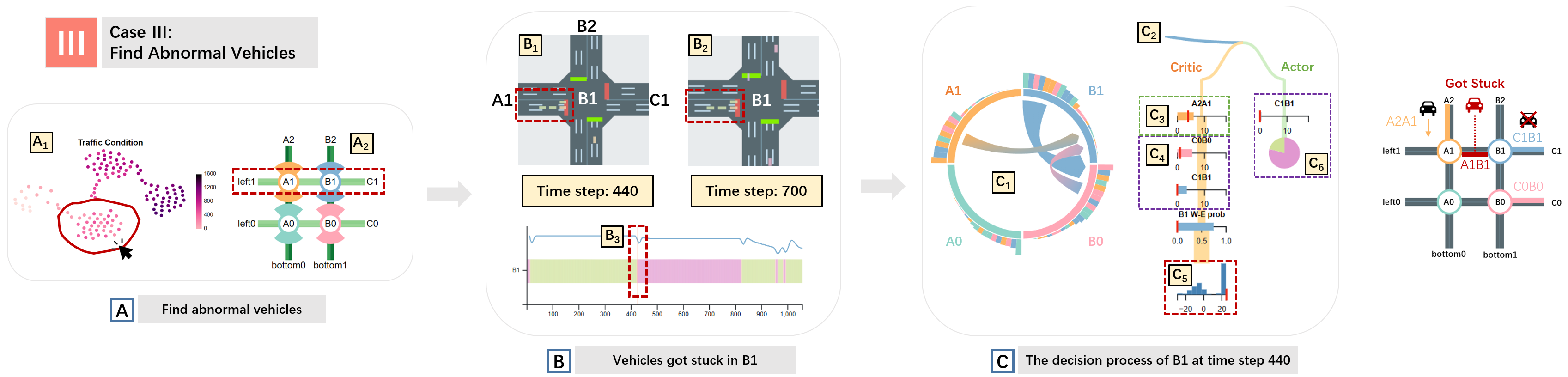}
    \vspace{-6mm}
  \caption{\revised{Experts' operations to explore abnormal vehicles. (A) Identify abnormal vehicles during exploration. (B) Vehicles got stuck in intersection B1 and the expert selected the time step when the agent changed the traffic signal. (C) The decision-making process of agent B1.}}
  \label{fig:case-3}
  \vspace{-3mm}
\end{figure*}

\subsubsection{\textbf{Case III: Anomaly Detection of the MARL model}}
\par \textbf{Identification of Abnormal Episode.} When E6 explored the training process in the \trainingdistributionCapital, he found an unusual episode $17$. As shown in~\cref{fig:teaser}B, the speed loss of episode $17$ was very low, but the reward has not reached the optimal level. He sorted episodes immediately according to speed loss in the \trainingdistribution~(\cref{fig:teaser}$\rm B_1$). He noticed that the speed loss of episode $17$ was the lowest in the training process. To understand what caused this phenomenon, he selected episode $17$ and explored it further in the \episodeoverviewCapital~(\cref{fig:teaser}C) and the \episodedetailCapital~(\cref{fig:teaser}D).

\par At first, for some agents, the size of sectors in the W-E direction is much larger than sectors in the N-S direction~(\cref{fig:teaser}$\rm C_2$), which means those agents keep the green light in the W-E direction on, even the main traffic flows are in the N-S direction. In the \episodedetailCapital, he focused on the metrics and traffic signal phases in~\cref{fig:teaser}D. As shown in the right part of \cref{fig:teaser}D, both agents A1 and B1 had a long green band. They all adopted the strategy of only allowing vehicles in W-E directions throughout the episode. Except for A0, the reward and critic value of the agents went down rapidly over time. Then, he checked the states of A1 and B1 in \cref{fig:teaser}$\rm D_2$, the projection was distributed in a chain. There were many light circles in \cref{fig:teaser}$\rm D_2$, which indicated low reward. Since the state's distribution of A1 and B1 were similar, E6 took A1 as an example for analysis. After zooming in, those light circles turned into glyphs in \cref{fig:teaser}$\rm D_3$. The large purple sectors in the N-S direction demonstrated there were more vehicles in the N-S direction. Considering the Shapley values in \cref{fig:teaser}$\rm D_1$, he realized that agents focused on its entry in the N-S direction (\textit{A2A1}). Combined with the {\simulationreplayCapital} (\cref{fig:teaser}$\rm F_2$), E6 noticed that the N-S direction was extremely congested because the W-E direction is always allowed to pass. Due to the congestion in the N-S direction, many vehicles were even blocked inside the intersection. This also caused traffic congestion in the W-E direction framed in \cref{fig:teaser}$\rm F_2$. He thought that the long-term congestion in the N-S direction greatly reduces the reward, making this N-S feature the most important one. To answer why the $17$ episode has the lowest speed loss with poor performance, he concluded, ``In this episode, most of the traffic flow in the W-E direction maintained smooth. The speed loss of them was 0 at most time steps. This greatly reduced the speed loss of the entire road network.'' He added, ``Compared with earlier episodes, the metrics of episode $17$ are better but its strategy is very extreme. In the past, we could only write programs to judge by ourselves, but now we can find out more quickly with the help of the system.''

\par \textbf{Find Abnormal Vehicles.} In the test episode, E6 wanted to explore how agents' strategy changes with the traffic condition according to the \episodeoverviewCapital. After he selected a time period in \cref{fig:case-3}$\rm A_1$, he found there were a few vehicles in the W-E direction of the road network~(highlighted by the red rectangle in \cref{fig:case-3}$\rm A_2$). ``At first, I thought these vehicles were remnants of the previous stages and would leave the road network soon, since the current stages only had traffic flows in the N-S direction.'' However, after he checked each intersection one by one. He found some vehicles got stuck in agent B1's intersection. Those abnormal vehicles are demonstrated in~\cref{fig:case-3}$\rm B_1$ and $\rm B_2$. Next, he used the {\episodedetailCapital} to select the time step when B1 changed its traffic signal phase (\cref{fig:case-3}$\rm B_3$). In the chord diagram in \cref{fig:case-3}$\rm C_1$, a chord from A1 to B1 and a chord from B0 to B1 show that B1's surrounding intersections A1 and B0 had an influence on its decision. Then he moved to the decision process with more detail in \cref{fig:case-3}$\rm C_2$. For the critic of B1 in \cref{fig:case-3}$\rm C_3$, he observed there were many vehicles in the N-S directions (\textit{A2A1}). In the W-E direction in \cref{fig:case-3}$\rm C_4$, he noticed two roads (\textit{C0B0} and \textit{C1B1}). Especially, \textit{C1B1} is one of the entries of B1's intersection and there was no vehicle on \textit{C1B1} at this time step. Although the critic of B1 got global information including one of its entries (\textit{C1B1}) as well as its surrounding intersections, it still ignored there were a few vehicles got stuck in \textit{A1B1} (\cref{fig:case-3}$\rm B_1$) and thought highly of B1's action in \cref{fig:case-3}$\rm C_5$. Similarly, for the actor of B1, it only focused on the vehicles on \textit{C1B1} to make the decision \cref{fig:case-3}$\rm C_6$. ``That's an interesting result, which shows why the model made a mistake at this time. One agent tends to pay more attention to the main traffic flows. As a result, vehicles in other traffic flows may be neglected. This may lead to unreasonable waiting time in the real world.'' To avoid this problem, E6 recommended adding more rules, such as considering the waiting time of the vehicle in the reward function or setting a time limit for each traffic signal phase according to traffic flow.

\par \revisedOld{To sum up, case III proves: (1) the limitations of judging the model training effects from certain metrics and the effectiveness of {\name} for identifying the abnormal situation. (2) {\name} has the potential to enable users to quickly find abnormal situations and explain why the model made the decision in a given state, which can provide suggestions for model improvement in the future.}

\subsection{Expert Interview}
\label{section:interview}
\par \revisedOld{We organized individual interviews with experts E1-E6 (their background has been introduced at the beginning of~\cref{section:expertPractices}). First, we briefly introduced our system and provided a simple tutorial to demonstrate the visual design and interactions of {\name}. Next, experts could explore the system for about an hour. Then we carried out an individual interview for each expert. Each interview lasted about 30 minutes and we collected their feedback as follows:}

\par \textbf{System Designs.}
\revisedOld{Generally, all experts agreed that our system is useful to explore and strengthen the interpretability of RL-based TSC problems. After the simple tutorial, they could easily understand the purpose and meaning of our visual designs. They thought our system provides enough information to track the RL model's training process and agents' behaviors. ``\textit{I can easily observe agents' reactions to different traffic conditions and directly compare how agents' strategies change over episodes.}'', said E3. Some experts mentioned that the existing tools, such as~\textit{TensorBoard}, cannot provide friendly support to combine traffic simulation platforms, which makes it hard to analyze RL models' decision-making process under a certain state. Besides, E1 said that ``\textit{sometimes we need to design a segmented reward function according to traffic condition. The Policy Explainer Module can provide suggestions on setting the threshold of segmentation.}''}

\par \textbf{Usability and Suggestions.}
Generally, experts approved that the modules of {\name} are user-friendly. It may be not easy to understand some visual designs at first glance, but after the simple tutorial, they could easily understand the purpose and meaning of our visual designs. Moreover, they agreed that {\name} provides a multi-level analysis for a MARL-based TSC model. ``\textit{It's insufficient to evaluate the model only based on metrics. {\name} is useful to understand the model more comprehensively.}'', said E5. Many experts approved that the findings at different levels are interesting, especially in exploring abnormal situations. For the \episodedetailCapital, E6 can easily get the main idea of the visual design of the {\episodedetailCapital}, but the middle part can be more intuitive (\cref{fig:teaser}$\rm D_2$). ``The state glyph can be turned into a real intersection.'' In addition, he said that we can add more connections between the {\episodedetailCapital} and the {\simulationreplayCapital}, such as showing reward in traffic simulation directly. E1 suggested that the system can take more factors into account.``\textit{The mixed traffic flow of different vehicles and traffic rules can affect how traffic conditions change.} Further, she valued the potential of the system to guide RL model design in the future. She said that ``\textit{The relation between agent and environment can guide us to design action and reward function. How the reward functions influence an agent's behaviors is worthy of further exploration.}''

\begin{figure*}[h]
  \centering
  \includegraphics[width=\textwidth]{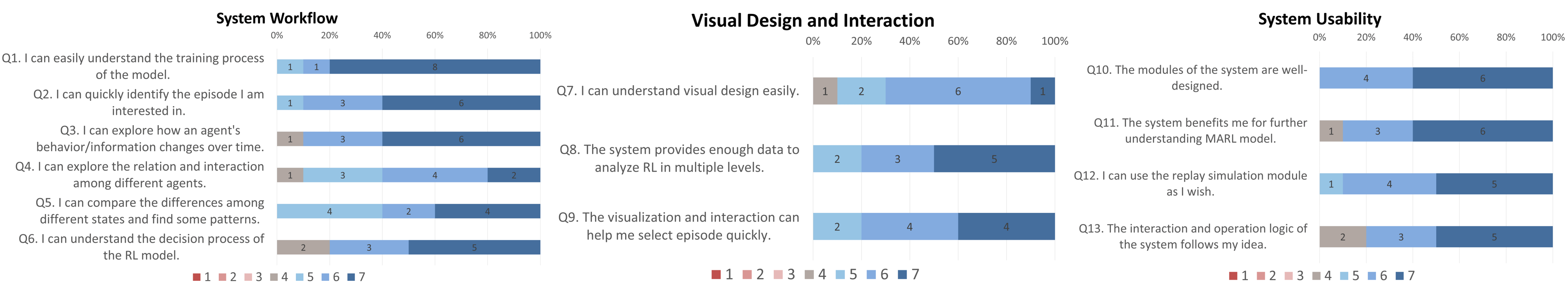}
    \vspace{-6mm}
  \caption{\revised{The result of questionnaires.} 1-7 represents ``strongly disagree'' to ``strongly agree'' for each statement.}
  \label{fig:userstudy}
  \vspace{-3mm}
\end{figure*}

\subsection{User Study}
\label{section:userStudy}
\par A user study is conducted to further evaluate {\name} in terms of \textit{system workflow}, \textit{visual design}, and \textit{system usability}.
\par \textbf{Participants.} We invited 10 participants ($3$ females, $7$ males, $age_{mean}$=24.1, $age_{std}$=1.59) majoring in traffic engineering from a local university. All participants possessed basic knowledge in TSC and RL. Specifically, 9 of them were master's students, while 1 participant was pursuing a doctoral degree. Their familiarity with RL and traffic simulation software, applied in their respective research areas such as TSC and autonomous driving, made them ideal candidates as target users for our system. The selection of these participants was strategic, as they could offer valuable insights into the practice of RL models. The user study was conducted in a face-to-face setting. We presented and demonstrated our system to the participants, who then actively engaged with the system to complete various tasks. Following the interactions, we utilized questionnaires to systematically gather feedback and suggestions from the participants, aiming to capture their perspectives on the usability and functionality of our system.

\par \textbf{Tasks.} In this user study, participants were tasked with completing three assignments. The outlined tasks are as follows: (1)\textbf{Task 1: Describe Model Training.} In this task, participants were instructed to articulate the trend of the training process based on metrics and evaluate the performance of the model. (2)\textbf{Task 2: Episode Exploration.} Participants were asked to select the last episode and describe the characteristics(action sequences, reactions to different traffic flows, states, et al.) and relationships among different agents. They were also encouraged to compare different episodes and synthesize observations on how agents evolve throughout episodes. (3)\textbf{Task 3: Understand Decision Process.} In this task, participants were tasked with choosing a state and elucidating the decision-making process of the RL model according to features and the value of the features.

\par \textbf{Procedure.} Initially, we provided participants with an overview of the traffic simulation settings, essential definitions of the example RL model, visual design components, workflow, and the system's overarching framework. Following this introduction, we conducted a comprehensive 20-minute demonstration of the system, delving into detailed explanations of each view's functions and visual design elements. Subsequently, participants had the opportunity to operate the system themselves and seek assistance from us as needed. Once participants felt comfortable, they were tasked with completing the aforementioned assignments using our system. Upon the completion of all tasks, participants were required to fill out a questionnaire featuring Likert scale questions. The Likert scale employed for the questionnaire ranges from 1 to 7, where 1 denotes ``strongly disagree'', and 7 indicates ``strongly agree''. 

\par \textbf{Results.} The results of the questionnaires are depicted in \cref{fig:userstudy}, indicating that participants held a favorable opinion of {\name}. Notably, the average score for each question surpassed 5.7. Regarding the system workflow, participants expressed satisfaction with their ability to comprehend the RL model's training process swiftly and efficiently select target episodes. The system was lauded for facilitating exploration of different states and comparisons among agents. In terms of visual design, the system was praised for furnishing ample information to support multi-level analysis, enabling participants to gain a comprehensive understanding of the RL model and identify patterns during training. Specific commendation was directed towards the design in \cref{fig:episode-detail-design-alternative}c. One participant remarked, ``\textit{Once I used the traditional diagram to compare traffic signal phases of different intersections in a road network. The diagram became too long at last, which made it hard to read and could easily lead to misunderstandings.} While participants generally found the system's modules well-organized for usability, there were suggestions for improvement. Some participants recommended providing an overview of each state cluster. In terms of system usability, participants broadly agreed that the different modules were well-organized, enhancing their understanding of the MARL model in the TSC scenario. Additionally, some participants expressed a desire for more typical TSC scenarios as examples and requested support for additional user-customized options.

\section{Discussion and Limitation}
\par \textbf{Improvements on the Current Workflow.} Based on the case study and interviews with domain experts, {\name} enhances researchers' workflow in three aspects. First, it offers a multi-level analysis of the model, which stands in contrast to the prevalent reliance on metrics and line charts for evaluating model performance. This expanded perspective allows researchers to comprehensively explore and analyze the model across varying time granularities, thereby enhancing their understanding. Second, {\name} significantly reduces the time required for researchers to unearth pertinent information. Handling high-dimensional data, such as states and traffic conditions, often presents challenges for researchers seeking to leverage them effectively for model exploration and evaluation. Researchers now can easily compare time series data generated by each agent within both the {\episodeoverviewCapital} and {\episodedetailCapital}. Last, the interpretability of the MARL model in the domain of TSC has received less attention in comparison to model performance. {\name} addresses this by striving to reveal the decision process of the MARL model, offering an intuitive means for researchers to comprehend the model and provide valuable guidance for future enhancements.

\par \textbf{Effective Communication with Domain Experts.} Despite the extensive research conducted by experts E1-E6 in RL-based TSC, their understanding of interpretability and visualization has been limited, leading to communication challenges. To address this, we have implemented strategies to enhance communication efficiency. First, in terms of visualization, experts typically utilized simplistic diagrams like line charts and traffic light phase diagrams during research, as depicted in \cref{fig:episode-detail-design-alternative}a. However, obtaining direct visual design requirements proved challenging. To overcome this, we introduced visualization cases from previous studies upfront, aiding experts in grasping the utility of visualization. Second, there existed a gap in the comprehension of interpretability. Initially, experts perceived interpretability as the integration of RL models with traditional theories in the TSC domain, lacking familiarity with commonly used model-agnostic techniques for explainable ML. Consequently, we provided an accessible introduction to interpretable techniques, with a primary focus on their practical usage. Last, in system development, experts faced challenges in specifying their visual design and interaction preferences. To address this issue, we adopted a rapid iteration approach, engaging in frequent communication with experts to optimize the system iteratively.

\par \textbf{Scalability and Generalizability.} Currently, our experiments employ a 2×2 grid network and a straightforward traffic flow setting, a configuration widely accepted in the traffic domain and endorsed by our domain experts. Realizing the necessity to incorporate more intersections (agents) for real-life applications, we acknowledge that while our existing designs can handle additional agents to some extent - such as the seamless addition of new agent information to the radical and top-down layout in the {\episodedetailCapital} - more effective strategies merit exploration. \revised{Consequently, given that {\name} is centered on TSC, its visual design could potentially be adapted for other traffic-related tasks. For example, the design in {\cref{fig:episode-detail-design-alternative}c} could facilitate the examination of traffic congestion and the coordination of different traffic signals. In terms of the utility of our system in other RL problems, the methods we used to interpret MARL models are model-agnostic, eliminating the need for users to integrate new structures into their MARL models. Although {\name} focuses on TSC, the tree-based design in the {\policyexplainerCapital} has less connection with TSC scenario, which can be used in other RL-based tasks.}

\par \revised{\textbf{Real-world Deployment of {\name}.} In contrast to computer vision domains, RL-based TSC models do not require extensive processing of complex visual data, thereby reducing the computational resource requirements significantly. Since training RL-based TSC models can be time-consuming, {\name} collects data throughout the model training process, eliminating the need for separate training or testing phases in our system. For the expert interviews and the user study, we have already deployed the system on various desktops and laptops, and generally, the overall operation is quite smooth.}

\par \textbf{\revised{Learning Curve and Evaluation.}} {\name} is designed for RL and TSC researchers aiming to enhance their understanding of MARL-based TSC. \revised{Through expert interviews and the user study, we dedicated more time to explaining the {\episodedetailCapital} and the {\policyexplainerCapital} due to their involvement with algorithmic interpretations of the model. Overall, users already familiar with RL-based TSC find it easier to grasp {\name}, while others may require a bit more time. Following a brief tutorial lasting approximately $20$ minutes, users can comprehend the design objectives of different components and believe that {\name} facilitates a more comprehensive understanding of the model. As for evaluating our system, existing tools commonly used by researchers, such as~\textit{CityFlow} and~\textit{SUMO}, primarily serve as programming frameworks for RL model development and traffic simulation. To the best of our knowledge, there is no directly comparable system to {\name}, which limits our ability to provide a more extensive evaluation of our system.}
\vspace{-1mm}
\section{Conclusion and Future Work}
In this study, we present {\name} to aid traffic and RL researchers in gaining a deeper, more comprehensible understanding of RL-based TSC issues. We identify a range of metrics for extracting key traffic information from the RL model. Our approach involves presenting and scrutinizing the RL model from various perspectives and dimensions, incorporating a traffic simulation module. We employed SHAP and decision tree techniques to demonstrate the relationships among different agents, shedding light on the decision process of the RL model. To assess the efficacy and user-friendliness of our system, we conducted three case studies, interviewed domain experts, and conducted a user study. The results validate that {\name} significantly enhances users' ability to explore the interpretability of RL-based TSC models. \revisedOld{Looking ahead, we plan to involve more end users and evaluate the long-term utility of {\name}. In addition, we aim to offer more customization options and add introduce a model comparison module to enhance direct comparisons between various RL-based TSC models.}




%

%
%
\vspace{-1mm}
\ifCLASSOPTIONcompsoc
  \section*{Acknowledgments}
\else
  \section*{Acknowledgment}
\fi
\textcolor{black}{
This work is supported by grants from the National Natural Science Foundation of China (No. 62302531) and the Science and Technology Planning Project of Guangdong Province (No. 2023B1212060029).}



\ifCLASSOPTIONcaptionsoff
  \newpage
\fi



\bibliographystyle{IEEEtran}
\bibliography{main.bib}

 \vspace{-10mm}

\begin{IEEEbiography}[{\includegraphics[width=1in,height=1in,clip,keepaspectratio]{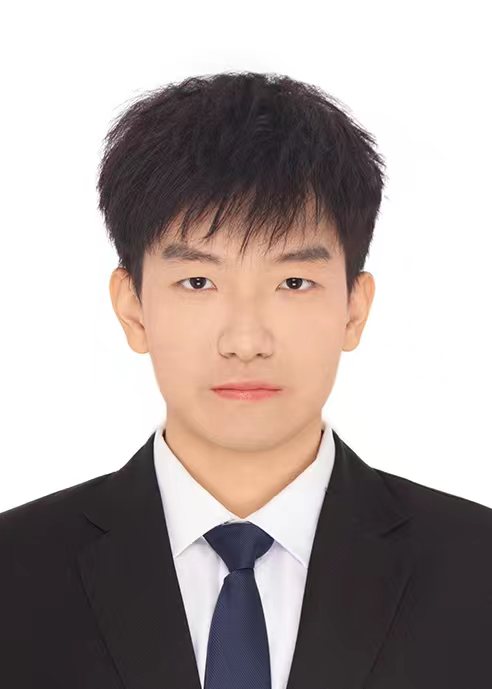}}]{Yutian Zhang} is currently working towards the M.S. degree in the School of Intelligent Systems Engineering at Sun Yat-sen University. He received a B.S. degree in transportation engineering from Sun Yat-Sen University. His research interests include interpretable machine learning, visual analytics and transportation big data.
\end{IEEEbiography}
 \vspace{-18mm}
\begin{IEEEbiography}[{\includegraphics[width=1in,height=1in,clip,keepaspectratio]{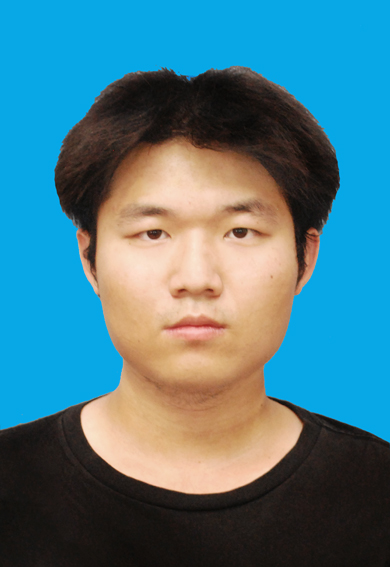}}]{Guohong Zheng} is currently working towards the M.S. degree in the School of Intelligent Systems Engineering at Sun Yat-sen University. He received a B.S. degree in transportation engineering from Sun Yat-Sen University. His research interests include interpretable reinforcement learning, visual analytics and transportation video analysis.
\end{IEEEbiography}
 \vspace{-16mm}

\begin{IEEEbiography}[{\includegraphics[width=1in,height=1in,clip,keepaspectratio]{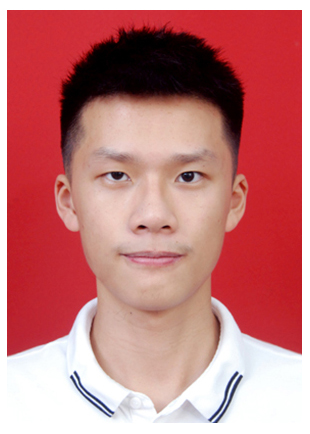}}]{Zhiyuan Liu}
is currently a senior undergraduate student in the School of Intelligent Systems Engineering at Sun Yat-sen University, majoring in transportation engineering. His research interests are in vehicle control, intelligent self-driving, and reinforcement learning.
\end{IEEEbiography}
 \vspace{-18mm}

\begin{IEEEbiography}[{\includegraphics[width=1in,height=1in,clip,keepaspectratio]{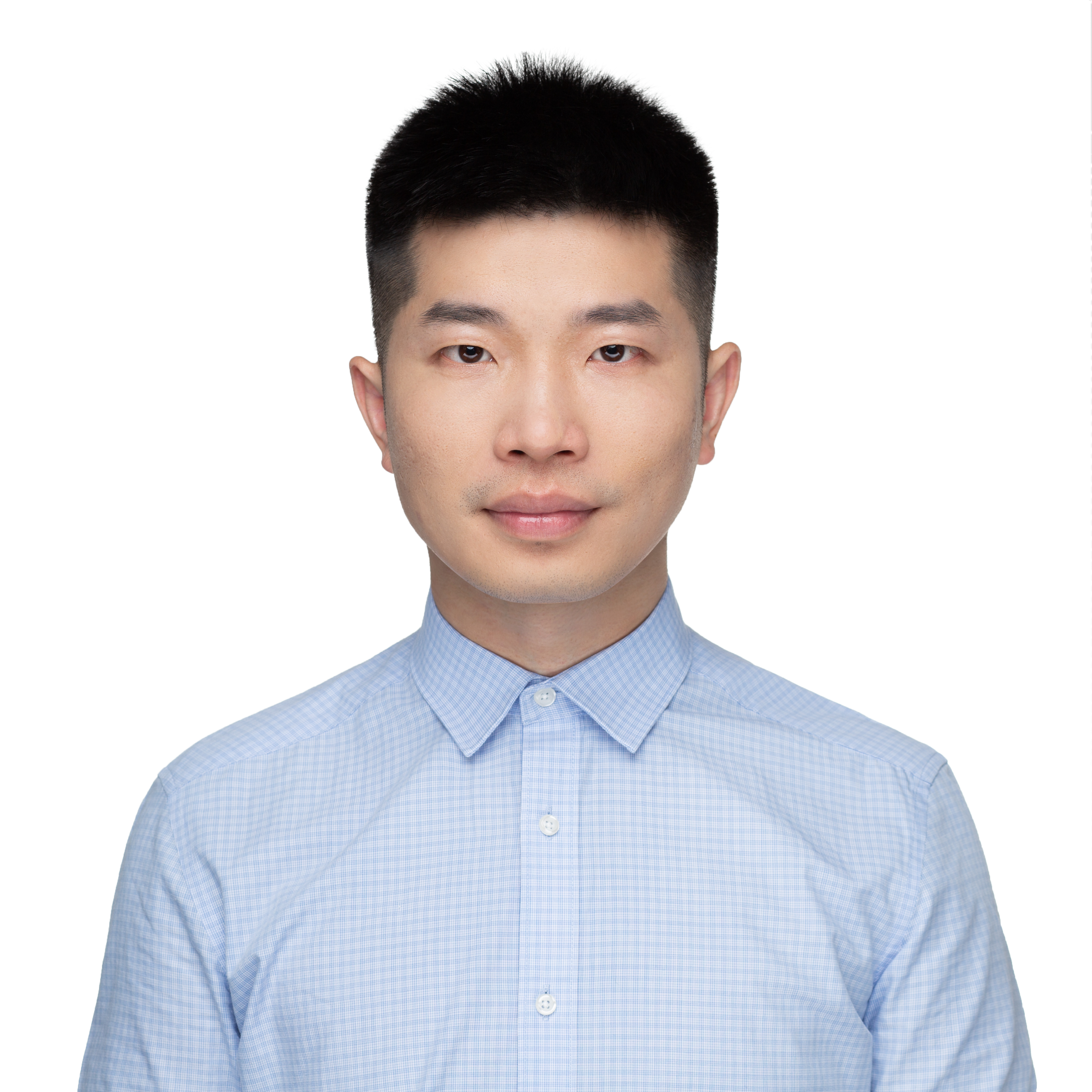}}]{Quan Li}
is currently a tenure-track assistant professor at School of Information Science and Technology, ShanghaiTech University. He received his Ph.D. from the Hong Kong University of Science and Technology. His research interests lie in visual analytics, interpretable machine learning, and human-computer interaction. For more details, please refer to \url{https://faculty.sist.shanghaitech.edu.cn/liquan/}.
\end{IEEEbiography}
 \vspace{-15mm}
\begin{IEEEbiography}[{\includegraphics[width=1in,height=1in,clip,keepaspectratio]{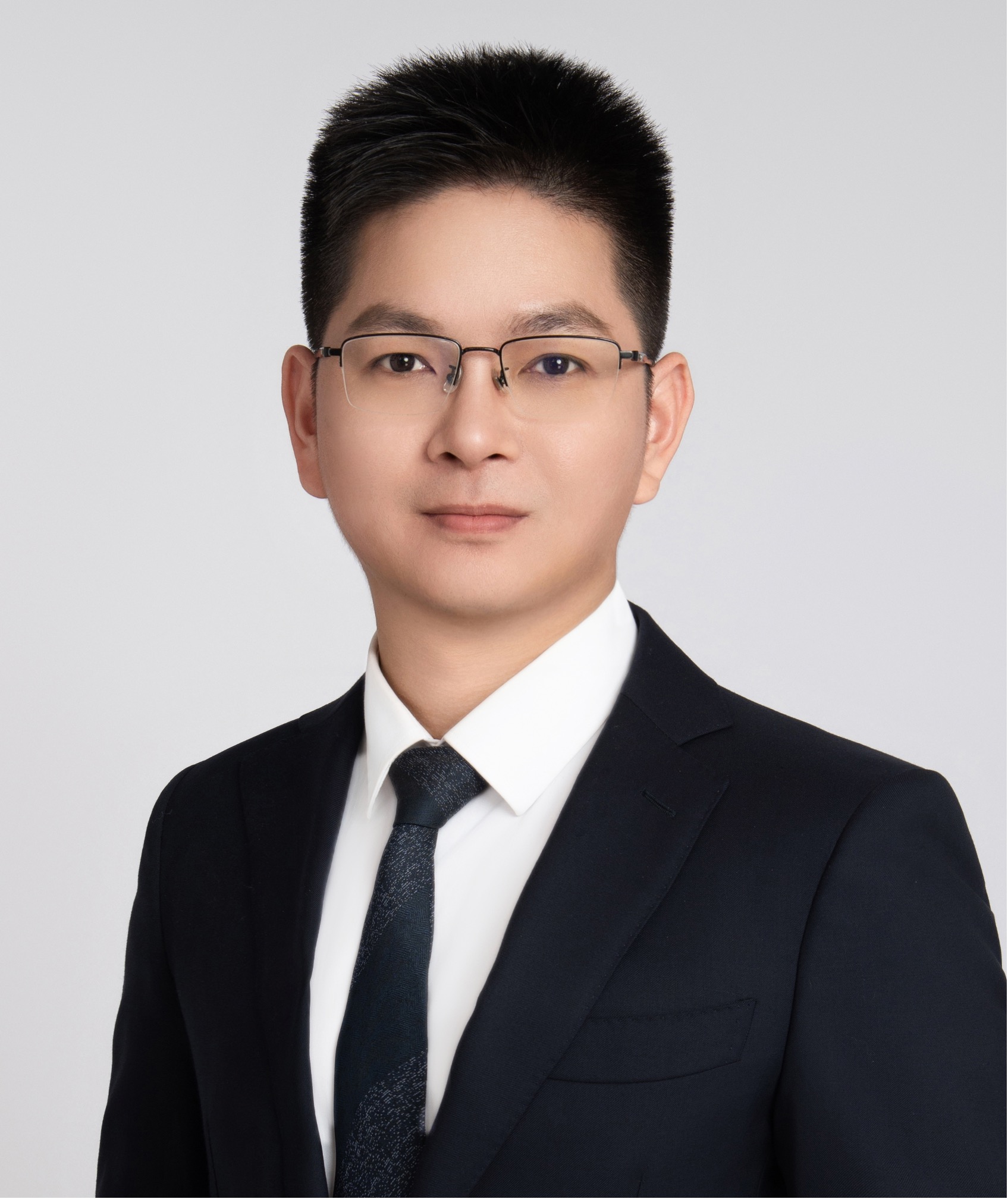}}]{Haipeng Zeng}
is currently an assistant professor in School of Intelligent Systems Engineering at Sun Yat-sen University (SYSU). He obtained his Ph.D. in Computer Science from the Hong Kong University of Science and Technology. His research interests include data visualization, visual analytics, machine learning and intelligent transportation. For more details, please refer to \url{http://www.zenghp.org/}.
\end{IEEEbiography}
 
\end{document}